\documentclass[letterpaper]{article} 
\usepackage{aaai2026}  
\usepackage{times}  
\usepackage{helvet}  
\usepackage{courier}  
\usepackage[hyphens]{url}  
\usepackage{graphicx} 
\usepackage{natbib}  
\usepackage{caption} 
\usepackage{algorithm}
\usepackage{algorithmic}

\usepackage{newfloat}
\usepackage{listings}
\DeclareCaptionStyle{ruled}{labelfont=normalfont,labelsep=colon,strut=off} 
\floatstyle{ruled}
\newfloat{listing}{tb}{lst}{}
\floatname{listing}{Listing}
\usepackage{hyperref}
\hypersetup{
    colorlinks=true,
    linkcolor=purple,
    filecolor=violet,      
    urlcolor=teal,
    citecolor=teal,
    pdftitle={Overleaf Example},
    pdfpagemode=FullScreen,
    }

\usepackage{makecell}
\usepackage{amsmath}
\usepackage{amsthm}
\usepackage{enumitem}
\usepackage{stmaryrd}

\usepackage{booktabs}
\usepackage{xspace}
\newcommand*{\eg}{e.g.\@\xspace}
\newcommand*{\ie}{i.e.\@\xspace}

\newcommand*{\cf}{cf.\@\xspace}

\newcommand*{\etc}{etc.\@\xspace}

\newcounter{rec}

\newcommand{\rec}{%
  \refstepcounter{rec}%
  \label{rec:\therec}%
  \textbf{R\therec}%
}
\newcommand{\rref}[1]{\textbf{R\ref{rec:#1}}}
\newcommand{\recommendation}[1]{\\[1mm](\rec)\ \textbf{\em #1}\\[1mm]} 

\theoremstyle{plain}
\newtheorem*{example*}{\textit{Example}}

\title{\emph{Law} of Large Numbers:\\ Accuracy as a Statistical Measure for AI Compliance and Competition}
\author {
    Rabanus Derr\textsuperscript{\rm 1, \rm 2},
    Alina Wernick\textsuperscript{\rm 1, \rm 3},
    Robert C. Williamson\textsuperscript{\rm 1, \rm 2}
}
\affiliations {
    \textsuperscript{\rm 1} University of Tübingen\\
    \textsuperscript{\rm 2} Tübingen AI Center\\
    \textsuperscript{\rm 3} CZS Institute for AI and Law\\
    rabanus.derr@uni-tuebingen.de, alina.wernick@uni-tuebingen.de, bob.williamson@uni-tuebingen.de
}

\begin{document}

\maketitle

\begin{abstract}
The machine learning community progresses (in part) by improving the 
  ``accuracy'' of its systems.
The  EU AI Act explicitly refers to ``accuracy'' as part of its compliance measures for high-risk AI systems.
  Are we talking about the same thing?
  This work presents ``accuracy'' as a case-study for differing requirements of social worlds, the technological machine learning community and the legal community.
  While competition on accuracy contributes to technological development, machine learning scholars simultaneously recognize accuracy's shortcomings regarding the usefulness and effectiveness of machine learning systems.
  The legal counterpart embraces the vagueness of ``accuracy,'' leaving interpretative flexibility for technological and societal changes. At the same time, accuracy is a core element of compliance within the EU AI Act.
  We elaborate on five main tensions, (a) nature of accuracy, (b) notion of performance, (c) scope of validity, (d) ends, and (e) statisticalness, to show that the two communities project disparate, and sometimes  contradictory, expectations on accuracy.
  Both legal and technical communities lack precise understanding of ``accuracy'' beyond the contextual boundaries of their community.
  The resulting frictions, \eg, based on the empirical or normative understanding of accuracy, are symptoms of an unresolved (and unresolvable) debate on what accuracy is.
  We constructively use the frictions to recommend baselines and interventional studies in standardization, and demand for tools to extend the validity of accuracy measurements.
\end{abstract}



\section[Introductions]{How to Read This 
Paper?}
This paper is structured unlike many others. There are two introductions. Plus, there are two threads in the main section of this paper (Section~\ref{L} and Section~\ref{E}). Are you a lawyer or rather from the social sciences? Then, take the left introduction on the following page. This introduction is marked with an ``L'' for ``law''. Are you a machine learning (ML) researcher or engineer? The right introduction on the following page is for you. You'll be guided by an ``E'' for engineer.\footnote{Such experimental approaches in scientific literature might be less common in machine learning. In legal research, there exist great examples of breaking the boundaries of standard scientific approaches, documents and formats, \eg, \citep{fuller1949case} and \citep{kang2004privacy}.}

\begin{figure}
    \centering
    \includegraphics[width=1\linewidth]{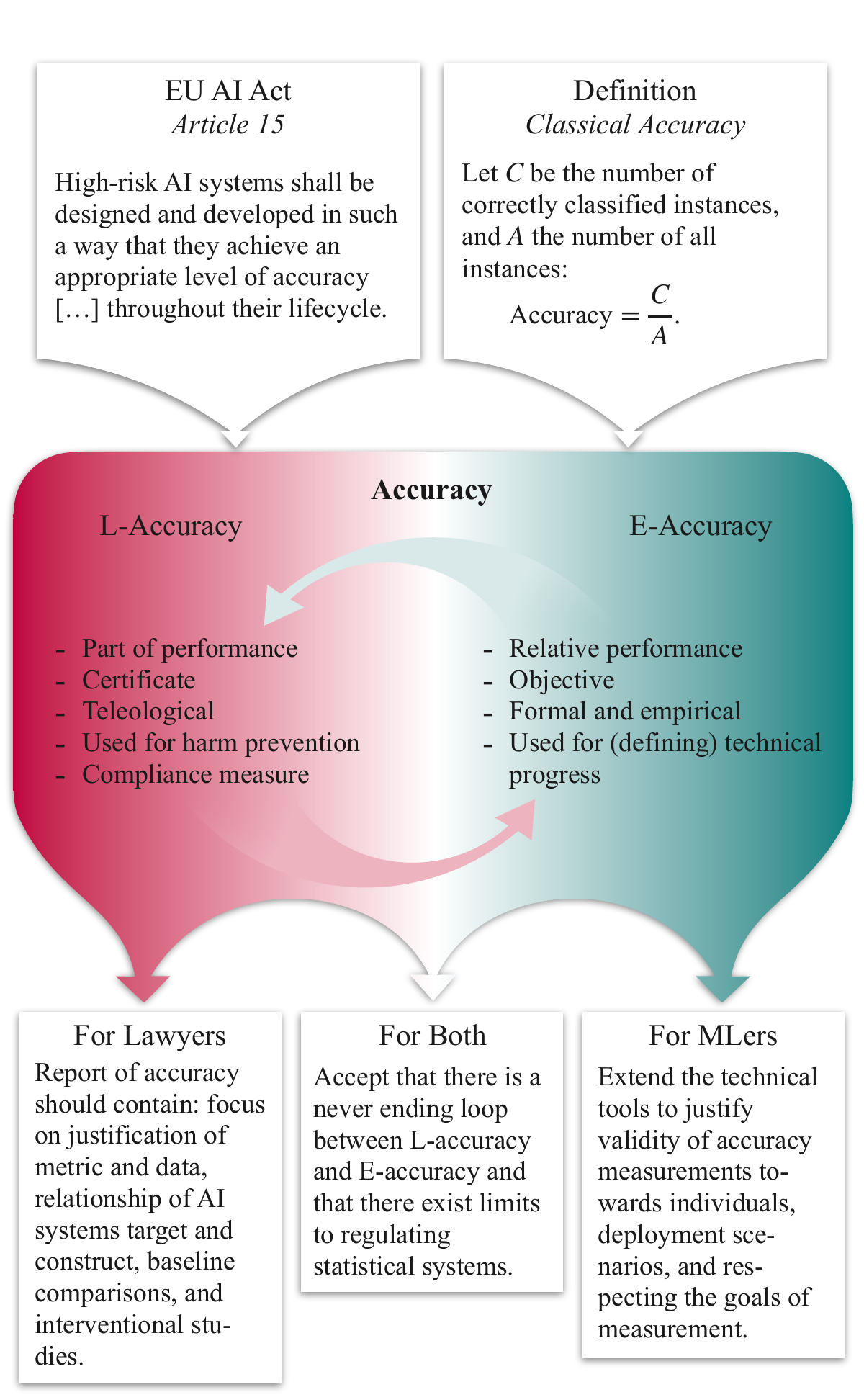}
    \caption{Graphical Summary of the Paper.}
    \label{fig:summary}
\end{figure}

\renewcommand{\thesubsection}{\arabic{section}.\arabic{subsection} L}
\subsection{Introduction for Lawyers}
In 2024, the European Parliament passed the EU Artificial Intelligence Act (AI Act, AIA). Enacted in response to a growing concern that AI ethics and self-regulation were insufficient for protecting against risks posed by AI \citep{smuha2025european,europeancommission2020ai}, it regulates the provision and deployment of AI systems in the EU. As a piece of EU technology law, the AI Act follows the blueprint of product safety regulation and aims at the harmonization of the internal market with the aim of supporting the adoption of trustworthy AI. Whereas the earlier product safety regulations sought to protect against risks to health and safety, the AI Act also addresses risks to fundamental rights \cite{almada2025eu}. 
The drafting of compliance standards for the EU AI act proved challenging. As a result, application of its obligations for high-risk systems, including accuracy, was postponed to 2027 and 2028\cite{Omnibus}.

Besides legal concepts, the AI Act also features terminology that originates from technical community \citep{koivisto2024user}. Some of them, such as ``explainability'' \citep{kaminski2021right} are explicitly normative. However, others such as ``robustness'' appear more technical \citep{nolte2025robustness}. One such central technical concept is ``accuracy''. Accuracy is part of the pivotal compliance obligations for high-risk AI systems (AIA Art. 15), \ie, AI application contexts particularly sensitive to risks to health, safety and fundamental rights, such as education and healthcare (AIA Art. 6; Annex III). For example, the provider of a high-risk AI system must ensure that the system performs with a consistent level of accuracy throughout the life-cycle (AIA Art. 15 (1); rec 73).

Furthermore, accuracy is a cornerstone of current advancements in machine learning \citep{Donoho2024Data, hardt2025emerging}. For engineers, accuracy is a (numerical) attribute of an \emph{algorithm} (a mechanical procedure) or a \emph{model} (essentially a complex mathematical structure) that is designed to make predictions, recommendations, or comparable outputs. Its ``accuracy'' is a mathematical formula (one of many possible choices) that evaluates the performance on particular test cases. 

The engineer strives to make an accurate algorithm in the same way she strives to make a fast car.

Given its centrality to both the legal and the technical community, one would expect ``accuracy'' to be defined within the EU AI act. It's \emph{not}. Hence, \textbf{what is accuracy?}




In this paper, we argue that a naive deferral of ``what is accuracy'' to the machine learning community or any self-contained doctrinal approach to the question is doomed to fail. The technical and legal community project different expectations on accuracy, use it in different ways, and follow different aims, \cf \citep{nolte2025robustness}. However, neither can claim to converge to a true single notion. Rather, the friction between the expectations reveal the constructive potential of ``accuracy''. In other words, we use ``accuracy'' as a case study for identifying translation gaps, misunderstandings and needs at the intersection of the technological world of machine learning and the social world of law.
\renewcommand{\thesubsection}{\arabic{section}.\arabic{subsection} E}
\subsection{Introduction for ML Scholars}
AI systems have left the laboratory and entered society. 
They have an impact on economic growth \citep{aghion2017artificial}. They transform jobs \citep{acemoglu2022artificial}. They raise (new) concerns about social fairness and the perpetuation of bias against marginalized groups \citep{barocas2016big,barocas2023fairness, hu2026new}. They prompt questions about legal accountability and the right to explanation in automated decision-making \citep{wachter2017counterfactual}, and they have led to dedicated regulatory frameworks \citep{eu_ai_act_2024}.
In these developments the machine learning community takes on a new relevance. It is not only the technology itself that is adopted and deployed, but also its terms, concepts, and assumptions.

A key concept and value(!?) \citep{birhane2022values} of machine learning is ``accuracy''. Accuracy underlies many recent advances via the wholesale adoption of benchmarking as the style of argument in machine learning \citep{Donoho2024Data, hardt2025emerging}. Originally, ``accuracy'' referred to the technical quality criterion of match-rate in classification, \ie, the relative number of correctly predicted classes. It's use, however, has overwritten that understanding with a more abstract interpretation as statistical comparison of machine learning system outputs with their desired counterparts.

Accuracy also is a central compliance measure within the EU AI act. For lawyers, accuracy is a quality certification that guarantees performance with respect to an intended purpose. The intended purpose is crucial: it makes no sense to regulate automobiles in terms of their use as medical interventions; their intended use is for personal transport. 
It explicitly takes into account the role of data generation, the importance of the transferability and robustness of accuracy to new application scenarios, and the harms and effects on society and individuals. Nevertheless, accuracy remains undefined and vague. The lawyer strives to guarantee accurate algorithms in the same way she strives to guarantee safe cars.

Given its centrality to both the legal and the technical community, one might expect that the machine learning community can ``simply'' explain to lawyers what accuracy is. It's \emph{not} that simple. Accuracy acquired new meanings, new facets and new expectations when it left the inner circle of machine learning. The community can not give a simple answer to the question: \textbf{what is accuracy?}

In this paper, we use ``accuracy'' as a case study for understanding interactions and frictions that arise when the technology of machine learning meets society. We claim that neither the technical nor the legal community have the ability to authorize a single universal notion of accuracy.
Instead, we highlight how accuracy is negotiated at the boundary. The friction between the expectations set on ``accuracy'', the use of ``accuracy'' and the aims of ``accuracy'' reveal a constructive potential, which leads to a set of recommendations to both the communities legal and technical. In other words, we use ``accuracy'' as a case study for identifying translation gaps, misunderstandings and needs at the intersection of the technological world of machine learning and the social world of law.

\renewcommand{\thesubsection}{\arabic{section}.\arabic{subsection}}
\subsection{Contributions}
Rather than considering what accuracy \emph{is}, we focus upon what accuracy \emph{does} in both  technical and  legal domains.
\begin{itemize}
    \item We identify \emph{E-accuracy} (Section~\ref{E}) as a statistical, formal and empirical way of defining the relative performance of a machine learning method in comparison to others. The comparison is used to compete for high accuracy.
    \item We elaborate \emph{L-accuracy} (Section~\ref{L}) as a performance compliance measure, with the normative aim to prevent harms. Validity concerns about accuracy turn out to outweigh concerns about the methods of measurement. We provide detailed doctrinal analysis in Appendix~L.
    \item The result of the parallel presentation is a structured comparison of the dimensions of friction (Section~\ref{consequences and conclusions of clashes and challenges}): (a) \emph{Nature of accuracy.} Is accuracy formal and empirical or rather normatively and teleologically oriented towards the intended purpose? (b) \emph{Notion of performance.} Does accuracy define performance or is part of it? (c) \emph{Scope of validity.} To which extent is validity necessary and possible, for comparison or for use-cases with risks of harm? (d) \emph{Ends.} Is accuracy used as an internal objective or rather an external certificate? (e) \emph{Statisticalness.} Is the focus of accuracy based on populations or individuals? 
    \item The named frictions offer ten, concrete, constructive recommendations, see \rref{1}--\rref{10}. These recommendations address the law community: We advocate for the reporting of accuracy with a focus on the justification of the use of metric and data, how the target of the AI system relates to the construct of intended purpose. The report, so we argue, should furthermore include baseline comparisons and interventional studies dependent on the context. 
    These recommendations as well address the machine learning community: There are new (technical) ways of ensuring the validity of accuracy measurement demanded which go beyond comparative performance. Furthermore, the machine learning community offers tools to approximate ``individual accuracy'' which can be helpful from a legal perspective. Finally, some of the recommendations address both communities: We need to accept there is no end to the sense-making loop when bridging formal and normative definitions of accuracy. We should be aware of the limits to regulating statistical systems which view the world through aggregates of quantities, and the side-effects of this statistical stance which is promoted in machine learning, and reified in the EU AI act. 
\end{itemize}

\section{Accuracy as a Boundary Object}
\label{method: accuracy as boundary object}
``Legal and political institutions lead, as much as they are led by, society’s investments in science and technology \cite{jasanoff2004ordering}.'' The technologies developed by AI-engineers are now used at scale in the wild. The intensified confrontation of the technology with society leads to a set of interactions in which society and societal norms are shaped and technical interests and requirements are redefined and co-produced \cite{jasanoff2004states}. This back and forth has entangled and enriched concepts such as ``accuracy''. To disentangle and ``squeeze the juice'' out of ``accuracy'', we treat ``accuracy'' as a \emph{boundary object}, shed light on it from two angles, and use it as a case-study to elaborate more generally on the interplay society and technology. In this section, we give background to our approach.

A \emph{boundary object} is an ``analytic concept of those scientific objects which both inhabit several intersecting social worlds [\ldots] and satisfy the informational requirements of each of them'' \cite{star1989institutional}. 
Boundary objects are subject to ``interpretative flexibility'' and thus diverging sense-making across different groups \cite{leigh2010not} --- in this case lawyers and engineers / scientists. Boundary objects function as an enabler for ``different groups to work together without a consensus'' \citep{leigh2010not}. We keep dissensus about the meaning of accuracy throughout the entire study.\footnote{This is a feature, not a bug.}

The boundary object ``accuracy'' has not emerged all of a sudden in-between AI and law. It has rather undergone a transformative development, being stretched by the expectations set on it from a variety of different angles. A stylized history of ``accuracy'' in machine learning and legal scholarships follows: We arbitrarily declare the starting point of ``accuracy'' with the adoption of ``classical accuracy'', \ie, rate of correct classifications, in machine learning, \eg, \citep{hughes1968mean, Duda1973-ed}.\footnote{The use of the term in this context likely goes back to the precision-accuracy divide known to measurement theory \citep{sep-measurement-science}. Simultaneously, the closely related receiver operator curve (ROC) gained renewed interest in signal detection theory \citep{peterson1954theory}. Even though ``accuracy'' was used within the context of classification, it was not simply defined as the mathematical definition Eq.~(1) we provide in the Appendix.} It initially appeared in the AI Ethics Guidelines of the EU Commission's High-level Expert Group of AI as a quality of a trustworthy AI \citep{HLEG}. However, it quickly morphed into the more sparsely worded legislative requirement in EU's White Paper on regulating AI \citep{europeancommission2020ai}. While included already in the first draft of the EU AI Act \citep{europeancommission2021aiact}, it lacks definition. It appears that the lawmakers referred to ``accuracy'' in the context of the AI Act, apparently presuming that the technological community knew what accuracy is. Subsequently, ``accuracy'' gained in relevance and developed to a more abstract term in the machine learning community, \eg, \citep{Gupta2020ResponseBT}. And, the circle continues.
Our precis is stylized, but shows the entanglement, mutual interaction, and the difficulty of separating a legal versus technical understanding of ``accuracy.'' We will cut the boundary object ``accuracy'' roughly following the line of machine learning and legal scholarship. We will use  \emph{E-accuracy} to refer to the technical, machine learning conceptualization, and \emph{L-accuracy} for the legal conceptualization. Whether the cut, which we propose, perfectly separates the understanding does not matter. 

\begin{figure}
    \centering
    \includegraphics[width=1\linewidth]{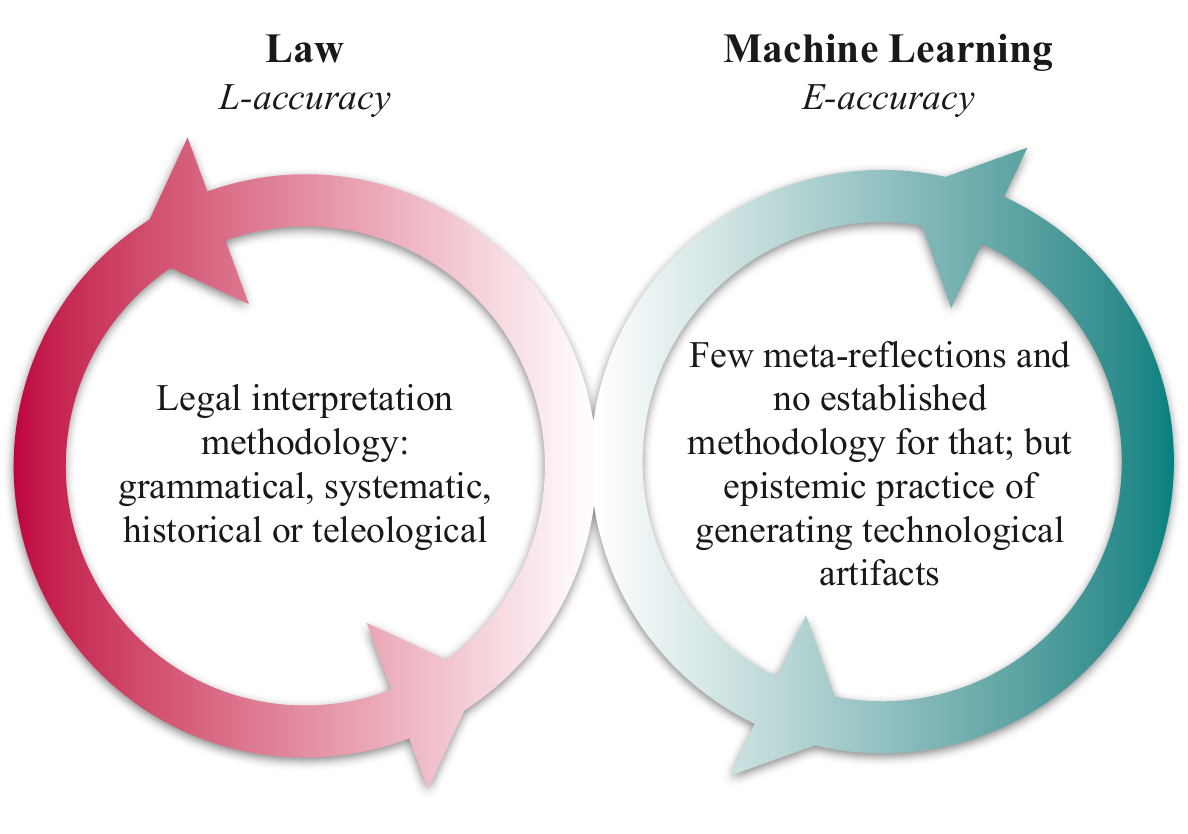}
    \caption{The stylized process of sense-making in the context of the term ``accuracy''. The conceptualization of the term ``accuracy'' in law (L-accuracy) and the conceptualization of the term ``accuracy'' in machine learning (E-accuracy) follow their own logics or dynamics. We are particularly interesting in the mutual interaction between the sense-making loops, which is caused by frictions in the understanding of the term ``accuracy'' (marked as white transition between the circles).}
    \label{fig:accuracy-sensemaking}
\end{figure}

We illustrate how the stylized origin story of ``accuracy'' is resolved in the constant dialogical sense-making and conceptualization between a technical and legal community in Figure~\ref{fig:accuracy-sensemaking}. External influences further shape ``accuracy''. These effects will be touched only marginally in this work.

The tensions arise because of the distinctive epistemic practices in both worlds. This becomes evident upon examining the question ``what is accuracy?''
Machine learning has no established methodology to answer that. Meta-reflections are largely absent within technical research, because they do not produce new technological artifacts. 
However, there exist commonly agreed upon epistemic practices for the development of technology, in which ``accuracy'' plays a vital role (Section~E.3). Hence, within this realm we elaborate on how accuracy is used, and approach a meta-reflection on accuracy by circulating around the technical literature. In particular, we mainly take into consideration existing accounts with a socio-technical focus (Section~\ref{E}).

Accuracy represents a technical concept introduced into law \citep{koivisto2024user}.  Such concepts, including ``algorithmic bias and fairness'' and ``explainability'' carry normative meanings also in the technical community, which may or may not find their correspondence in law \citep{kaminski2021right,weerts2023algorithmic}. For example, explainability techniques are viewed insufficient to meet the normative standards that the law sets, or should set for algorithmic decision-making \cite{edwards2017slave, cofone2026institutional}.

Technical terms represent linguistic transplants to law. When left undefined, they create uncertainty about their interpretation \citep{koivisto2024user}. One the one hand, the technical community may give multiple meanings to the given concept \citep{weerts2023algorithmic,koivisto2024user}. On the other hand, when integrated into law, the concepts become subject to the practices of legal interpretation \citep{koivisto2024user}. The requirement for ``accuracy'' in the AI Act, must be seen through the regulatory objectives set for the AI Act. Similarly to compliance criteria present in other product safety regulations, the conditions for high-risk AI are reflective of a metaregulatory approach. The EU legislator enacts relatively abstract criteria for high-risk AI, which are to be specified by harmonized standards set by standard-setting organizations CEN and CENELEC \citep{EC2025C3871}. In the absence of, or instead of, the more detailed standards, the providers may choose their own way to comply with the requirements \citep{aiact2025standards}.

This means providers -- \ie, computer scientists -- interpret and give meaning to the notion accuracy in the routine compliance with the AI Act. However, when incorporated into legal act, the term accuracy becomes part of law \citep{koivisto2024user}. Lawyers and the judiciary will define the meaning of the term based on established methods of legal interpretation. Common methods of interpretation are (a) grammatical (focus on the wording), (b) systematic (focus on the system of rules, \ie, the AI Act and other EU law), (c) historical (preparatory works; earlier related pieces of law), (d) teleological (focus on the objectives of the AI Act and its impacts) \citep{mollers2022legal}. Critically, this legal maneuver might uproot terms from one habitat, and appropriate them to law -- with unpredictable consequences for future legal interpretations \citep{koivisto2024user}.


Through E-accuracy and L-accuracy we will observe two epistemic disciplines with self-reproduction of their knowledge \cite{luhmann1992wissenschaft}. 
Both disciplines make sense of ``accuracy'' in their own terms. For instance, machine learning accepts ``accuracy'' only to be expressible in formal, technical language, otherwise, it is not of value for the development of technological artifacts (Section~E.2). On the other hand, legal scholars seek for consistency of the use of ``accuracy'' across legal text. That results in an understanding of ``accuracy'' as a normative, teleological concept (Section~L.4). The surprising observation is that both sides can easily refer to their sense-making as the ``right'' and ``true'' one, a phenomenon called ``self-vindication'' by Ian Hacking \citep{hacking1992self}: It is truth what we found, because we used a valid method; the method is valid because it leads to the truth. In this text, we accept both the ``truths'', the ``true'' meaning of accuracy as E-accuracy and the ``true'' meaning of accuracy as L-accuracy.

\subsection{Organization of the Paper}
The remainder of the text is structured closely to the methodological approach laid out above.
Below we present E-accuracy and L-accuracy in parallel threads in a double-column format (Section~\ref{L} and Section~\ref{E}). In the main paper we summarize the claims. In the supplementary material (Appendix~L and Appendix~E of the full version \citep{derr2026law}), we give the ``proofs'' for those claims.
Both threads are independently readable, but mutually complementary. 
The tensions created in these sections are then resolved, evolved and convolved in Section~\ref{consequences and conclusions of clashes and challenges}.
The two ``truths about accuracy'' reveal requirements for action, \eg, in standardization and technical research, and fundamental limits in the reconciliation process. We conclude by taking a satellite-perspective on the relationship between machine learning and law.

    \renewcommand{\thesection}{\arabic{section} L}
\section{What is L-Accuracy?\\ AI Lawyer's View on Accuracy}
\label{L}
Being the first comprehensive effort to regulate AI, the EU AI act aims to improve the functioning of the internal EU market and promote the uptake of human-centric and trustworthy AI (AIA rec. (1)). 
A key compliance requirement within the Act is ``accuracy'', a concept which is alluded to also in the context of documentation. Despite its prominent role, ``accuracy'' is not defined. 
We outline below how and in which context ``accuracy'' is used within the EU AI act.

\subsection*{Introduction to \textcolor{teal}{regulation} of (computational and statistical) \textcolor{purple}{products} (L.1)}
2024 the EU AI act entered into force. The EU AI act is a product safety regulation concerned with AI systems. It is based on risk classifications which depend on the use of the AI system. In particular, high-risk AI systems, \eg, when used in healthcare or education, need to fulfill a series of compliance requirements.

\subsection*{L-accuracy is a \textcolor{teal}{legal compliance} measure (L.2)}
Among other compliance measures, high-risk AI systems need to ``achieve an appropriate level of accuracy'' following Article 15 of EU AI Act. More specifications and demand on the reporting of accuracy are scattered in the Act. In fact, ``accuracy'' is not explicitly defined. However, it can be considered as part of ``performance'', the ability of an AI system to meet its intended purpose.

\subsection*{L-accuracy \textcolor{teal}{certifies} the \textcolor{purple}{fitness-to-purpose} of machine learning \textcolor{violet}{systems} (L.3)}
The EU AI act aims to certify the development process of AI systems. Part of this certification is accuracy. In particular, accuracy is to be understood with respect to the intended purpose of the AI system. The intended purpose, within the scope of high-risk use cases, can range from predicting risk of school drop-out to interventional scenarios such as allocating support based on drop-out risk.


\subsection*{L-accuracy is inherently \textcolor{teal}{normative} and used for \textcolor{purple}{risk mitigation} (L.4)}
With the requirements around accuracy, the EU AI act aims to mitigate risks for health, safety and fundamental rights. Risk mitigation, so it is argued, is achieved by ensuring accuracy ``throughout the lifecycle''. That reading makes L-accuracy \emph{inherently} normative. In the anticipating attempt to reconcile the normative with a technical understanding, the AI Act highlights, beyond the transparency of the accuracy measurement, the centrality of validity (arguments) of the accuracy measurement.


\subsection*{L-accuracy \textcolor{teal}{exists to protect} subgroups and individuals  (L.5)}
Generally, law has a focus on individuals and individual's rights. More specifically, accuracy is explicitly mentioned in the context of subgroups and individuals within the EU AI act. Accuracy can serve to protect them.


\addtocounter{section}{-1}
\renewcommand{\thesection}{\arabic{section} E}
\section{What is E-Accuracy?\\ ML Engineer's View on Accuracy}
\label{E}

Accuracy is so central and standard to machine learning that the question what it \textit{is} feels inappropriate. Nevertheless, carving out the facets of E-accuracy will serve several purposes here.
We make explicit what is unarticulated, implicit knowledge in machine learning. In doing so, we emphasize aspects which might seem marginal to the machine learning community, which, however, turn out to be fundamental anchors for tensions with respect to L-accuracy.

\subsection*{Introduction to \textcolor{teal}{engineering} of (computational and statistical) \textcolor{purple}{technologies} (E.1)}
Much of machine learning has become engineering, that is the construction and maintenance of a technology. In particular, machine learning creates computational systems for generating predictions, recommendations, or other artifacts. Critically, those systems are statistical, and hence are based on aggregates of data.

\subsection*{E-accuracy is a \textcolor{teal}{formal and empirical} measure (E.2)}
Classically, accuracy is understood as the match-rate between predicted and actual classes. More generally, accuracy involves two main ingredients: metric and data. Accuracy is the formal comparison (based on the metric) of desired and generated outputs, \eg, classes, rankings, regression values, on an aggregate of instances (based on the data).
Hence, accuracy is outcome-focused, formal, empirical and statistical.

\subsection*{E-accuracy \textcolor{teal}{defines} the \textcolor{purple}{performance} of machine learning \textcolor{violet}{models} (E.3)}
Accuracy is used as an abstract performance objective and even defines performance. The closer to deployment, the more the limits of accuracy are acknowledged even though accuracy still gives decision support on which model to use. Missing validity, \eg, because of mismatches in data, metric, constructs and interventions, limits the scope of accuracy in defining model performance.

\subsection*{E-accuracy is inherently \textcolor{teal}{statistical} and used for \textcolor{purple}{ranking models} (E.4)}
Accuracy is \emph{inherently} statistical (beyond E.2), as it favors statistical methods over other approaches, because those methods are optimized for exactly that goal (E.3). Currently, this optimization progress is measured in \emph{benchmarks}, a type of standardized accuracy measurement, which is based on ranking models. The use of accuracy in machine learning (benchmarking) only requires such relative comparisons. The validity of such rankings beyond the measurement scenarios are backed, but hard to hold beyond comparisons.

\subsection*{E-accuracy \textcolor{teal}{is poorly suited to} subgroups and individuals (E.5)}
Individual accuracy, \ie, the claim that a statistical system is accurate on a single, unobserved instance, is a hardly justified and even oxymoronic concept.
Statistical systems, more generally, act on the level of aggregates.


\renewcommand{\thesection}{\arabic{section}}
\section{From Clashes to Recommendations}
\label{consequences and conclusions of clashes and challenges}

In order to understand and mediate the resulting tensions of AI in socio-technical reality, we picked out ``accuracy'' as a recurring topos in legal text on AI, as well as scientific and engineering literature. 
We now elaborate on clashes of expectations and conceptualizations at the boundary. Despite the destructive connotation of ``clash'', we are convinced that the themes of friction and disagreement are productive starting points for both communities. In particular, we reveal the necessity for technical progress, recommendations for the ongoing standard-setting process for accuracy \cite{StandardCenelec} and fundamental limits to technical developments and remaining challenges which ask for changes in law, marked as recommendations.
The structure of this section is given in Table~\ref{tab:clash of E-accuracy and L-accuracy}. To facilitate understanding we introduce a running example.
\begin{example*}
    Imagine an automated AI-tool ``E-Grad'' for grading English exams in high-schools. This tool likely classifies as high-risk AI system under the EU AI act. Hence, it needs to fulfil the specified demands on accuracy. Furthermore, there is little additional and intertwining regulation relevant to educational scenarios different to healthcare.
\end{example*}
Our recommendations, especially \textbf{(R4),} \textbf{(R6)} and \textbf{(R9)}, target the ongoing standardization process for high-risk AI. To our knowledge, the issues highlighted have not been visibly addressed in the standardization process, which is at an early stage and opaque. The standards are prepared by CEN/CENELEC Technical Committee JTC 21. The sub-project for drafting of the accuracy standard (prEN 18229-4) began in June 2026 within its Working Group 4 on Foundational and ethical aspects of AI \cite{StandardCenelecdate,StandardCenelec, CENCENELECJTC21}. In addition, compliance criteria for accuracy are likely to be informed by standards for computer vision (prEN 18281) and natural language processing (prEN ISO/IEC CD 23282), the revised ISO/IEC DIS 4213 standard for AI performance evaluation and the standards for other high-risk AI obligations, such as transparency \cite{AI-sandards}, see also \cite{marin2026your}. 

 %

\begin{table*}
\small 
\centering
\begin{tabular}{l>{\raggedright}p{2.9cm}>{\raggedright}p{4.7cm}>{\raggedright}p{4.7cm}p{2.5cm}}
Sec. & Aspect                              & L-Accuracy                               & E-Accuracy           & Recommendations                                                                          \\ \midrule\addlinespace
\ref{clash:nature of accuracy} & Nature of accuracy                  & Teleological (Towards intended purpose)                     & Formal and empirical & \rref{1}                                                                          \\ \addlinespace
\ref{clash:notion of performance} & Notion of Performance               & Part of performance for intended purpose          & Defines comparative performance          &  \rref{2}, \rref{3}, \rref{4}, \rref{5}                                             \\ \addlinespace
\ref{clash:scope of validity} & Scope of Validity                   & For mitigation of risks in actual deployments                           & For comparison         & \rref{6}, \rref{7}                                                          \\ \addlinespace
\ref{clash:ends} & Ends                                & External certificate (to get better social outcomes)       & Internal objective (to improve and choose method) & \rref{8}\\ \addlinespace
\ref{clash:statisticalness} & Statisticalness                     & Focus on individual stance & Population based statistical stance & \rref{9}, \rref{10}                                                      \\ \addlinespace
\end{tabular}
\caption{Aspects along which E-accuracy and L-accuracy clash.}
\label{tab:clash of E-accuracy and L-accuracy}
\end{table*}

\subsection{Nature of Accuracy}
\label{clash:nature of accuracy}
While E-accuracy is a formal and empirical concept, which can  be written down mathematically and computed on data (E.2), L-accuracy remains elusive with its aim of use within the EU AI act as a centrally defining feature (L.4).
As the EU AI act acts on actual technology, a translation of the formal, empirical concept E-accuracy into its teleological counterpart is necessary. \citet{marin2026your} elaborate extensively on the choice of accuracy metric, the appropriate threshold and the test data. That entire process can be understood as the attempt to equip formal definitions of accuracy a teleological meaning. That particularly requires the effort for proper justifications of the choices made (Section~\ref{clash:scope of validity}).

The reconciliation of the natures of accuracy is (successfully) underway, \eg, standardization \citep{EC2025C3871}, new forms of evaluations like arenas\footnote{\url{https://arena.ai/leaderboard}}, and research developments \citep{marin2026your}. Without questioning the progress, we want to emphasize a lesson of the confrontation of E-accuracy and L-accuracy. The different natures of accuracy, which are tightly linked to the languages of the communities, are in principle \emph{not} translatable. L-accuracy defeats a full formalization and expression as an empirical measurement, because by its teleological definition it needs to follow the vague and mobile aims, such as as mitigating risks (Section~L.4). Conversely,   E-accuracy represents a static, momentary status. That discordant note, however, need not concern us. \recommendation{We have to accept that the search for formal and empirical representations of accuracy within the context of legal compliance will continue indefinitely.} For the better, that way accuracy respects the dynamics of our society and values and the friction motivates the development of new technical measures. 



\subsection{Notion of Performance}
\label{clash:notion of performance}

From the perspective of L-accuracy, ``performance'' is an overarching term, to which accuracy, but as well robustness or cybersecurity, belongs to (L.2). In particular, performance, including accuracy, needs to be considered with respect to the intended purpose of the AI system (L.3).
That holistic reading of performance and its pragmatic reading of accuracy, \ie, accuracy with respect to the intended purpose, is reflected in the limitations of E-accuracy (E.3).
While we argued that E-accuracy defines performance, it does so only within a limited scope. Critically, E-accurate machine learning models do not translate into successful applications for the same reasons for which E-accuracy does not translate into L-accuracy. The mismatches given in (E.3) can be understood as translation difficulties from E-accuracy to L-accuracy.

Accuracy metric mismatch is the problem that the choice of accuracy metric needs to capture performance semantics. Hence, \recommendation{a report of an accuracy metric needs to argue why it is semantically meaningful.} Data set mismatch is the problem that the choice of data on which accuracy is measured needs to be ``representative'' and meaningful for the deployment context. That is, \recommendation{a report of a data set needs to argue why the accuracy measurement on this data set is valid in a relevant scope, \ie, ``throughout the lifecycle'', and respects robustness in deployment changes.} Target-construct mismatch is the problem that the target which the AI systems aims to predict, recommend, generate is only in weak relationship to the abstract construct of intended purpose. It follows that \recommendation{a report of accuracy needs to specify the AI system's target and its relationship to the intended purpose.} The named mismatches are in fact concerns of missing validity (\cf Section~\ref{clash:scope of validity}), see \citep{marin2026your} for an elaboration on the first two challenges. 
\begin{example*}
    Accuracy of E-Grad measured as agreement of grade by tool and teacher(s) sounds reasonable. \rref{2}: But why does agreement matter? It reflects teacher's gradings, but shouldn't it rather matter that all potential spelling errors are found and considered for the grade? \rref{3}: Who is the teacher with whom the tool is compared with? From the same school? Averaged over multiple teachers from multiple schools? \rref{4}: Is the intended purpose of the AI system to reduce the workload of the teacher while the English skills of the students remain well-captured and why does the grading matter then?
\end{example*}

\paragraph{A Plea For Decision-Accuracy}
\label{clash:a plea for decision accuracy}
The EU AI act implicitly assumes that accuracy is a performance measure which applies to decisions produced by AI systems (L.2). In addition, the EU AI act does not make any restriction on whether the intended purpose of an AI system is interventional or not, \ie, whether the success of the AI system deployment needs to take into account changes happening due to the deployment (L.3).
\begin{example*}
    If E-Grad's grades are used to decide which student to support by an exchange program, the agreement of grading between tool and teacher are less relevant than the success of assigning students to the exchange program. In this case, what matters substantially are the effects on the English skills and whether the ``correct'' group of people was sent to the program.
\end{example*}
We suggest to make an interpretative move towards \emph{decision-accuracy} when there is an interventional purpose of an AI system.
That is, we ask for accuracy to be understood as measuring the accurate assignment of treatment with maximal causal effect. That requires a lot of the machinery of quantitative social science: (a) specification of the desired causal effect, (b) randomized data for evaluation, and (c) a model of effect \citep{angrist2009mostly}.
The standard understanding of accuracy in machine learning is far from any such interpretation and practice.
Nevertheless, under certain circumstances the standard understanding of accuracy is a good proxy for decision-accuracy \citep{pearl2021causally,imbens2021breiman, fernandez2022causal}.\footnote{Do we need counterfactual predictions of outcomes, \eg, potential outcomes \citep{rubin1974estimating} or counterfactual loss \citep{koch2025statistical}, and randomized (interventional) data to circumvent the problems mentioned?
Partially. First, it is clear that causal effect estimation, or  counterfactual predictions of outcomes, is sufficient for ``accurately assigning treatments to maximize causal eﬀect'' \citep{fernandez2022causal}. However, it is not necessary.
There are conditions under which accurate outcome predictions (prediction-accuracy) can be useful for interventions: (a) when outcomes and effects are correlated, (b) when the gain in estimation accuracy is strong in comparison to identifying the best treatment (c) when predictions need rank consistency but not absolute accuracy \citep{fernandez2022causal}.}
However, if those conditions are not met, E-accuracy becomes a meaningless measure for the performance with respect to an interventional intended purpose \citep{barabas2018interventions}.
The harmonization legislation of certain high-risk AI systems (Annex I(a)) already require interventional studies: MDR requires the proof of causally verifiable clinical benefit (Art. 2 (44)(52)). In comparison to \cite{eu_mdr_2017}, EU AI act's notion of ``performance'' makes not reference to causality (Art. 3(18). However, the overarching objective of risk-mitigation and risk profiles of Annex III (AI applications) support interpreting the requirement of accuracy to include ``decision accuracy'' and interventional studies. It should be acknowledged that interventional studies may require real-world testing within (Art. 57) or outside AI regulatory sandboxes (Art. 60). 
\recommendation{Standardization for the EU AI act needs to address accurate treatment assignment with causal effects.} Tools such as randomized control trials, differences-in-differences, \etc \citep{huntington2021effect, angrist2009mostly} should be part of the default canon of accuracy measurement techniques.
From a technical side, we'd profit from a clearer understanding of the proxy relationship between E-accuracy and decision-accuracy, \eg, \citep{fischer2025value, fischer2026empirically} and how accurate treatment assignments can be efficiently established \citep{casacuberta2026good}.

\subsection{Scope of Validity}
\label{clash:scope of validity}
For E-accuracy, the scope of validity is set by its use as a competition marker (E.4). Valid relative comparison of methods needs  to be possible beyond the accuracy measurement setup. In contrast, L-accuracy is rather more focused on its scope of validity ``throughout the[ir] lifecycle'' (AIA Art. 15(1)) and under ``known and foreseeable circumstances'' (AIA Art. 13(3)(b)(ii)) for mitigating risks in actual deployment (L.4).
There is a mismatch in the scope of validity, which let's us join the chorus of criticism regarding current AI evaluations, \eg, \citep{hutchinson2022evaluation, eriksson2025trust}.

Judgements about ``goodness'' of AI systems require contextualization.
It is certainly true that with benchmarks we observe intradisciplinary progress and that
``machine learning runs on a weaker currency: comparisons and rankings'' \citep{hardt2025emerging}.
But, concretely, for compliance to the EU AI act, or more abstractly, for following normative social requirements, {machine learning scholarship needs to serve the interests outside the discipline for extended scopes of validity of accuracy measurements with a focus on harmlessness and social benefit.} More concretely, the mismatch in the scope of validity brings the following two aspects to light.


\paragraph{A Request for Baseline Reportings in Legal Compliance to Extend Validity and Demonstrate Alternatives}
Machine learning scholarship observed that comparative accuracy measurements are substantially better justified to be valid beyond the measurement scenario than direct scores (Section~E.4). In other words, model rankings defined by their accuracy often transfer reliably to new test scenarios, while the actual accuracy scores do not. Thus we advocate
\recommendation{report  baselines and comparison systems as part of the standards for EU AI act compliance.} Baselines can extend the validity of accuracy measurements.


Furthermore, baselines demonstrate alternatives to the use of an AI system.
In \citep{wang2024against} the authors advocate that an AI system needs to prove its advantages along a set of dimensions over existing methods to justify its use.
For instance, in \citep{dressel2018accuracy} the authors show that non-experts and extremely simple statistical methods perform as well as a presumably complicated recidivism score prediction tool COMPAS. The strand of literature on the ``simpler'' alternative, \eg, following \citep{letham2015interpretable, semenova2022existence}, substantiates this approach.
Hence, reporting the baseline is crucial as it enables deployers and users to take a substantiated choice of whether and how to use an AI system.

\begin{example*}
    Accuracy reporting for E-Grad via a benchmark which consists of annotated, graded English exams at UK high-schools should be accompanied by baselines. That is, the report should contain benchmark results of, \eg, comparable AI systems or selected teacher re-gradings to substantiate the accuracy claims.
\end{example*}

The introduction of comparison systems comes with the burden to choose a comparison system. Which comparison system should we consider? Rule-based systems \citep{johnson2022bureaucratic}? Simple statistical baselines \citep{letham2015interpretable}? Humans? Competitor systems?
The provider needs to take the burden of defining and justifying a reasonable comparison system, which at least is (a) relevant for the intended purpose, (b) has shown accuracy achievements and (c) is distinct to the process of data set creation. Those conditions seem trivial, but can, depending on context, exclude certain choices. For instance, humans are often inaccurate \citep{dawes1989clinical} and not optimized for certain abstract accuracy metrics \citep{recht2025actuary}, plus they are often central players in the data creation pipeline. For instance, if the labels for images were given by certain human annotators, one needs a specific argument why a comparison of a machine learning system should be pursued against a new set of human labelers.
Hence, it is easy to exclude the possibility that there is a naive, universal choice of comparison system. Rather, the choice of a comparison system, as its justification, is bound to the context and purpose of the AI system.

\paragraph{AI Systems Lack a Diverse Set of Certifications}
On a more abstract level, the challenge of validity of performance conditions such as ``accuracy'' is not specific to AI systems. For instance, proving a drug's effectiveness by clinical trials is often based on statistical tests which are claimed to be valid beyond the participating cohort.
That inductive step from trial cohort to unknown patient is held to be justified, \eg, because the used randomized control trial safe-guards against hidden confounders.
Beyond the statistical test, the drug's effectiveness is certified, \eg, by bio-chemical tests and knowledge of effect mechanisms and the structure of the drug. Those certifications are even legally mandated, \eg, \citep[Article 8(3)(i)]{directive2001_83_ec}. With the diverse arsenal of methods, the scope of validity can be refined, \eg, if the drug's mechanism is known to rely on certain enzymatic complexes which diabetics don't share, then the drug is known to be useless for such patients.
Hence, in which way is the validity challenge for accuracy of AI systems special?

The validity challenge is particularly difficult for AI systems, because (a) AI systems are often treated as a general-purpose technology\footnote{The EU AI act is neutral to the set of (intended) purposes (AIA Art. 2, Art. 3(1)) and even defines ``general-purpose AI models'' and ``general-purpose AI systems'' (AIA Art. 2(63)(66)). The theoretically-inclined machine learning scholars make no semantic commitments regarding the inputs and outputs of their treated systems. The business-oriented community (over)claims no limits to the use-cases of machine learning products \citep{widder2024watching}.} and (b) AI systems of current interest are statistical.

Regarding (a) the variety of use-cases demand a diverse set of certifications of validity which needs to be represented in standards and regulation.\footnote{There is always some margin for refinement left here for any product safety regulation. For instance, what is the exact use of a medical device or a construction product and the relevant ways how they perform? However, the scope is still much narrower as such products are already defined through their abstract purpose, \eg, \citep[Art. 2(1)]{eu_mdr_2017} and \citep[Art. 3(1)]{eu_cpr_2011}.} Regarding (b), the statisticalness is not only a property of the AI system, but as well a central component of the problem formulation.\footnote{In fact, the EU Council wrote in the version of the EU AI Act from November 25 2022:
``The range of problems addressed by machine learning typically involves tasks for which other approaches fail, either because there is no suitable formalisation of the problem, or because the resolution of the problem is intractable with non-learning approaches'' \citep{CouncilEU2022AIAct}.} It follows that the compliance of such a statistical system needs to heavily   rely on statistical measures, such as accuracy. There is no causal, mechanistic, social, biological, chemical, physical, etc. way to argue that a certain AI system is accurate.
Drug effectiveness, in contrast, can be corroborated, \eg, by bio-chemical knowledge; there is another means for refining the scope of validity. \recommendation{Machine learning systems, hence AI systems, lack a diverse set of such certifications, beware of that.}

\begin{example*}
    E-Grad as AI-tool based on machine learning methods is a statistical technology. Patterns of text (English exam) to number (grade) are distilled from training data sets. Roughly, the patterns are distilled based on empirically testing a set of possible mappings and choosing the best among them. The choice of tested mappings is rather broad and general, often large neural networks. There exist little to no arguments beyond ``the chosen mapping is the statistical best'' which can be used to corroborate that E-Grad as a tool will work as expected under varying conditions.
\end{example*}


\subsection{Ends}
\label{clash:ends}
E-accuracy is used as an \emph{internal} objective within the machine learning community to improve and choose methods (E.3 \& E.4). L-accuracy certifies \emph{externally} the quality to users, deployers and other affected persons (L.3 \& L.4).
Despite the distinct uses of accuracy, ``being accurate'' is positively connoted. Everybody wants accurate AI systems, but in which way?

The ``iron rule'' of machine learning following \citet{hardt2025emerging} is the competition on benchmarks, an institutionalized measurement of accuracy. In its modern use, there are almost no rules on how to compete, except maybe for training or memorizing new methods on test sets \citep{xu2024benchmarking}.
Hence, ``gaming'' the competition is what happens (\cf \citep{blum2015ladder}), but is part of the definition of progress (\cf \citep{orr2024ai}) and is hence construed as a desirable feature.

Very differently, accuracy as compliance measure is a burden to the provider. There is a necessity for being accurate which is (a) not fully intrinsically motivated, but rather implied by the regulator, and (b) might not be aligned with respect to one's own intentions. There is likely an interest by the legal stakeholders to be accurate, but to demonstrate harmlessness, certain accuracy measurement might be needed which are against the interest.
A simple example is the classification accuracy of a diagnosis tool: provider, deployer and patients might agree that a low false negative rate is desirable, \ie, not many actually sick patients are diagnosed as healthy. However, a low false positive rate might not be in the interest of the deployer, hence of no interest to the provider, who will sell more treatments, but in the interest of patients who do not want to undergo unnecessary treatment.
So, providers can safeguard themselves against the accuracy compliance measure by carefully designing an irrelevant but easily fulfilled accuracy metric.
\recommendation{We need to anticipate that ``accuracy-hacking'' will matter and develop remedies against it }(\cf fairness-hacking \citep{black2024d, simson2024one, meding2024fairness}).

\subsection{Statisticalness}
\label{clash:statisticalness}
In contrast to E-accuracy and machine learning methods, which are statistical and thus focused on the aggregate, L-accuracy refers to individuals (L.5). This tension is particularly problematic, as we argued that from the perspective of E-accuracy ``individual accuracy'' is oxymoronic (E.5).
\begin{example*}
    If E-Grad has $95\%$ agreement with a teacher's grades on a set of English exams, this accuracy report does not and cannot tell anything about whether one individual exam (even in test set) was accurately graded. What can be said about the accuracy of E-Grad on an unobserved English exam?
\end{example*}

\paragraph{Practically Reconciling the Statistical and the Individual Stance}
A legitimate step forward is to interpolate between individuals and the overall aggregate. The EU AI act explicitly mentions not only individuals but as well ``subgroups'' (Section~L.5). In that regard ``individuals'' are the limit of a spectrum of finer and finer subgroups. Hence, while ``why is the AI system accurate on this individual'' is hardly justified (E.5), the question ``why is the AI system accurate on this subgroup'' at least partially relieves the burden of the validity argument.

In this regard, it might be helpful to demand for instance-wise reports of AI system's outputs and their desired counterparts, \eg, along the line of \citep{burnell2023rethink}. That way, subgrouping along different post-hoc specified categories is possible. Together with tools from fair machine learning, which often are concerned with worst-case performance on subgroups \citep{zafar2017fairness, pleiss2017fairness, hebert2018multicalibration,donini2018empirical, williamson2019fairness, loi2022calibration}, and the analysis of Rashomon sets, \ie, sets of machine learning models with equal performance \citep{frohnapfel2026using},\footnote{Concretely, the authors of \citep{frohnapfel2026using} suggest to study the disagreement of machine learning models. Disagreement can be an indicator for the invalidity of of an individual accuracy measurement beyond the measurement context.} valid approximations of ``individual accuracy'' will be practically possible. Hence, by refraining from the ideal of individual accuracy, \recommendation{we should use implementable methods to obtain justified estimates for subgroup accuracy or indication of failure of individual accuracy.}



\paragraph{EU AI act Codifies Statisticalness of Our (Social) World}
Besides the challenges around individualization, the EU AI act is an example of codifying a statistical stance into law, which itself usually is closer to an individual stance.
Due to the demand of accuracy, which has attached a statistical interpretation, statistical methods such as machine learning techniques gain a new type of foothold on a social level. The development-internal justification of machine learning systems, \ie, good statistical accuracy, attain a new relevance beyond the developing community.
The EU AI act hence reproduces \emph{patterns of ``performance''} shaped by engineers and developers.
Yes, the EU AI act can\emph{not} be ignorant towards the technology it wants to regulate. Nevertheless, it performs and reifies ``accuracy'' as a central goal for the technology of machine learning \citep[\S 2.1.2]{hildebrandt2020law}\citep{maccormick1986institutional}. In that regard, one might consider it a literal \emph{law} of large numbers.

The problem is that the patterns of ``performance'' might not be congruent with a broader societal consensus.
From a machine learning perspective, measuring everything in accuracy aligns well to its own principles. From a broader perspective, the relevance of accuracy (Section~E.3) in particular, and the representation of the (social) world in statistical terms in general, can be questioned.
For instance, \citet{eaglin2019technologically} argues that statistical systems for risk assessment in criminal justice restrict our ways of talking about justice.
It is hard to imagine what the exact consequences of the choice of ``accuracy'' as central statistical compliance measure within the EU AI act are.
But, \recommendation{we should consider the side-effects of a statistical stance to the world, promoted by ML and legislation.}
\begin{example*}
    Imagine the EU AI act legitimates the use of the AI-tool E-Grad. Hence, it promotes that statistical grading systems are accepted in our society. Do we want grading to be procedure governed by quantified data comparison? And how does that relate to the educational purpose of grading in the first place?
\end{example*}






\section{Lessons Learned -- A Satellite Perspective}
We examined two parallel threads of sense-making to ``what is accuracy'' from a machine learning and a law perspective in the above. Actually, the first meta-lesson we learned is the productive step from ``accuracy'' as a noun to ``accuracy'' as a verb. We more productively asked ``what accuracy \emph{does}.''

Both perspectives project their own needs, expectations and limits on accuracy, but agree that there is some concept of interest.
This pattern is not specific to ``accuracy'' at all.
Terms and concepts such as ``robustness'' \citep{freiesleben2023beyond, nolte2025robustness}, ``explanation'' \citep{bordt2022post, bordt2025position}, ``interpretability'' \citep{molnar2020general}, ``privacy'' \citep{bailie2025topics}, ``fairness'' \citep{barocas2023fairness} are created and negotiated at the boundaries of machine learning. The self-referencing circles ``what is ...'' within the technosphere break apart the more the technology leaves the laboratory.
Our study of ``accuracy'' holds some general lessons for all those terms and concepts.

\textbf{The clash of non-overlapping interpretations of technical terms is inevitable.} We even should not expect those clashes to settle (\cf Section~\ref{clash:nature of accuracy}).
Machine learning is restricted by its need for formalization and mathematization. 
However, that creates a force which easily clashes with other (social) perspectives. For instance, law is restricted by its attempt to consider competing social interests which often results in ambiguity and vagueness. Ambiguity and vagueness are the enemies of proper formalization (\cf debates on fairness metrics, \eg, \citep{barocas2023fairness}).

A second conflict arises along the line of statisticalness. Machine learning is based on statistics, while legal or other perspectives often are concerned with the individual. The statistical stance versus the individual stance will productively generate disagreement about interpretation (\cf see the debate on explanations, \eg, \citep{bordt2022post, bordt2025position}).  Relatedly, while there is plenty of precedent in regulating new technologies, there is not much precedent in the regulation of \emph{statistical} technologies. It now seems that it is not just philosophers but also lawyers who have to grapple with the question of how does the aggregate relate to the individual.

Finally, the internal logics of machine learning, the drive towards new methods \citep{sambasivan2021everyone, birhane2022values}, the solutionism \citep{reyes2025resisting}, oppose the internal logics of other social spheres such as law, with an ideal of consistency \citep[\S 2.2.2]{hildebrandt2020law}. Unsurprisingly, non-engineering ways of thinking don't need technical solutions for self-vindication. Still, the ways of thinking interact, and making sense of technical terms inevitably ends in a clash.

Even though clashes (will) occur, they (will be) are productive. The disagreement on meanings projects inside machine learning as well as outside. Exemplarily, the use of ``fairness'' got refined in machine learning \citep{barocas2023fairness} and sparked a debate beyond the technical community. \textbf{The clash of non-overlapping interpretations is productive and refines our understanding of the term.} Furthermore, such clashes repeatedly teach us the lesson to not put too much trust in abstract nouns with a naive belief in univocity.\footnote{The difficulties that abstract nouns create is nicely demonstrated by \citet{williams2014keywords} who grapples with examples such as ``equality,'' ``individual,'' ``masses,'' ``society,'' and ``technology.''}

What remains open as an interesting and important avenue for future work, is the study of power structures in defining relevant terms and interpretations \cite{jasanoff2004states, koivisto2024user}. How are frictions utilized, by whom for which purpose? Near term, it is relevant to study whether terminologies adopting existing standards influence the notion of accuracy in the EU AI act's standard setting process \cite{ISOStandardfunctionalcorrectness, Smithdeng}, and how the JTC 21 working groups' modular approach standard drafting for AI act Arts. 9-15 aligns with the holistic nature of L-Accuracy cf.\cite{marin2026your, AIA15Hart, derr2026law}.





 \section*{Acknowledgments}
 Thanks to Mara Seyfert for discussions;
 James Bailie, 
Riikka Koulu and Suvi Sankari for literature suggestions;
and Emeline Banzuzi for feedback. This work was supported
by the Carl Zeiss 
Foundation,  the Kone Foundation 
(project “Long-term human rights risks 
of smart city technologies”) (AW); 
the Deutsche
Forschungsgemeinschaft (EXC number 2064/1;  
Project number 390727645) (AW, RW) and the T\"{u}bingen AI Center (RD, RW). 



\appendix


\renewcommand{\thesection}{L}
\section{What is L-Accuracy? AI Lawyer's View on Accuracy}
\label{L:What is L-accuracy? -- Appendix}
Representing the first comprehensive effort to regulate AI, the EU AI act aims to improve the functioning of the internal EU market and promote the uptake of human-centric and trustworthy AI (rec. (1)). 
A key compliance requirement within the Act is ``accuracy'', a concept which is alluded to as well in the context providers' obligations of documentation. Despite its prominent role, ``accuracy'' is not defined within the EU AI act. 
Hence, in this section we outline how and in which context ``accuracy'' is used within the EU AI act. We try to sketch L-accuracy for a technical community. We show that:
\begin{enumerate}
    \item Accuracy is a part of compliance measures and performance of (high-risk) AI systems.
    \item Accuracy certifies the fitness-to-purpose of AI systems.
    \item Accuracy serves the purpose to mitigate risks.
    \item Accuracy is inherently a normative concept.
    \item Accuracy exists to protect subgroups and individuals.
\end{enumerate}

\renewcommand{\thesubsection}{L.\arabic{subsection}}
\subsection{Introduction to \textcolor{teal}{regulation} of (computational and statistical) \textcolor{purple}{products}}
\label{L:CONTEXT}


\paragraph{What is the EU AI act?}

 The AI Act represents EU product safety regulation (rec. 9). Following the so-called New Legislative Framework, the EU seeks to harmonize the inner market by establishing criteria for ensuring that the regulated products are not harmful to health or safety \citep{EU2008R765, EU2008D768, EU2019R1020}. The New Legislative Framework is characterized by EU regulators' reliance on harmonized standards to specify the compliance criteria. The product manufacturers can ensure legal compliance by complying with the key standards before bringing their products on the market \citep{blueguide, mueck2025european}. 

\paragraph{What is the goal of the EU AI act?}

EU AI act aims to improve the functioning of the internal EU market and promote the uptake of human-centric and trustworthy AI (rec. 1, Art. 1(1)).  To this end, the Act aims, on the one hand, to ensure a high level of protection of health, safety, fundamental rights, including democracy, the rule of law and environmental protection, against the harmful effects of AI systems in the Union and, on the other hand, to support innovation (Art. 1 (1)). 

\paragraph{To the end of trustworthy AI, the Act reflects a risk-based approach to regulating AI.}
The strictness of rules applicable to AI systems are reflective of the scope and intensity of risks generated (rec. 26). The Act's risk categorization is probablistic, rather than context-driven. It reflects the lawmaker's ex ante presumption of risks associated with specific use purposes \citep{almada2025eu}. It outright prohibits systems posing unacceptable levels of risk, such as subliminal techniques, social scoring systems and certain applications of emotion recognition, biometric recognition and categorization systems (Art. 5). High-risk systems are subject to most detailed regulation. Such systems can be brought to the EU market, provided that their providers meet the relevant compliance criteria (Arts. 3 (3), 8; 16).
High-risk systems encompass AI applications in areas where it may pose particular risks to health, safety or fundamental rights: biometrics; critical infrastructure; education and vocational training; employment, workers' management and access to self-employment; access to and enjoyment of essential private services and essential public services and benefits; law enforcement; migration, asylum and border control management and administration of justice and democratic processes (Art. 6(2); Annex III). Moreover, an AI system is classified as high risk, when it is intended to be used as\textit{ a safety component of a product, or the AI system is itself a product} covered by the Union harmonization legislation, where that legislation requires the product to undergo a conformity assessment procedure by a third-party conformity assessment body. Such products encompass for example medical devices, \textit{in vitro} diagnostic medical devices, lifts and safety components for lifts, toys and machinery (Art. 6 (1); Annex I (A); rec. 50). The definition and scope of the term "safety component" was specified by AI Omnibus \cite{Omnibus} (Arts. 6 1(a)-(c)).Whether or not an AI system is a high-risk system depends on the use for which an AI system is intended by its provider, also referred to as the ``intended purpose'' (Art. 3(12)). In addition, the EU AI act imposes transparency measures to certain AI systems, such as systems interacting with natural persons (Art. 50). Finally, the Act established compliance rules for General Purpose AI Systems (GPAI) as well as stricter criteria for GPAI systems with systemic risk (Chapter V). 

Since the compliance obligation for accuracy only concerns high-risk AI systems (Art. 15, Section), subsequent mentions of AI systems presuppose the system to be high-risk.

The compliance of the EU AI act apply to AI \emph{systems}. It is noteworthy that the term is distinct from the concept of AI \emph{model}, more commonly used in machine learning \citep{nolte2025robustness}. AI systems encompass more components, \eg, hardware, pre-processing or postprocessing procedures of inputs and outputs to the AI system, and particularly they do not necessarily refer to systems following the machine learning engineering practices \citep{rossi2025ai}. In this text, we presume a strong relationship between ``AI systems'' and ``machine learning models''. 


\subsection{L-accuracy is a \textcolor{teal}{legal compliance} measure}
\label{L:IS}

\paragraph{Compliance Measures for high-risk AI systems in EU AI Act}

The compliance obligations for high-risk AI systems are central to the fulfillment of Act's objectives. The majority of the obligations concern the providers of high-risk AI systems. A \emph{provider} is a natural or legal person, public authority, agency or other body that develops an AI system places it on the market or puts the AI system into service under its own name or trademark, whether for payment or free of charge (Art. 3(3)). However, the EU AI act also poses obligations to other actors on the AI value chain, most importantly deployers. \emph{Deployer} refers to ``a natural or legal person, public authority, agency or other body using an AI system under its authority'' (Art. 3(4).) 
  
The key compliance requirements for the providers of high-risk AI systems are contained in the Chapter 3, Section 2 of the EU AI act. They include the establishment of a risk management system (Art. 9 ); data governance and management practices (Art. 10); technical documentation (Art. 11 ); transparency and provision of information to deployers (Art. 13); Human oversight (Art. 14 ) and finally, the requirement of accuracy, robustness and cybersecurity (Art. 15 ). In addition, AI provider are also subject to  other obligations, such  as the requirement to establish a establish a quality management system (Arts. 16, 17). 

The breach of compliance obligations for high-risk AI systems are punishable by a fine up to 15 million euros or a maximum of $3\%$ of global turnover, whichever is higher (Arts. 99 (4); 101 (1)). Initially, the compliance obligations of high-risk AI systems were supposed to come into force, and become enforceable in 2026 (Art. 113, Chapters III and V). However, due to the lack of availability of harmonized standards, the EU has postponed the entry into application of rules on high-risk AI systems to December 2027 (Annex III systems) and August 2028 (Annex I systems)  \cite{Omnibus}. 

\paragraph{Compliance Measure of Accuracy and Context}
\label{para:compliance measure of accuracy and context}
As explained in Section 2, we focus on the criterion of accuracy (Chapter III, Section 2, Art. 15), because it represents a technical term, which is left undefined in the AI Act. 

Article 15 is entitled ``Accuracy, robustness and cybersercurity''. It requires that
\begin{enumerate}
    \item High-risk AI systems shall be designed and developed in such a way that they achieve an appropriate level of accuracy, robustness, and cybersecurity, and that they perform consistently in those respects throughout their lifecycle.
    \item To address the technical aspects of how to measure the appropriate levels of accuracy and robustness set out in paragraph 1 and any other relevant performance metrics, the Commission shall, in cooperation with relevant stakeholders and organisations such as metrology and benchmarking authorities, encourage, as appropriate, the development of benchmarks and measurement methodologies.
    \item The levels of accuracy and the relevant accuracy metrics of high-risk AI systems shall be declared in the accompanying instructions of use.
    \end{enumerate}

The recital 74 further specifies that ``high-risk AI systems should perform consistently throughout their lifecycle and meet an appropriate level of accuracy, robustness and cybersecurity, in light of their intended purpose and in accordance with the generally acknowledged state of the art''.

Beyond Art. 15, explicit and implicit accuracy-related compliance requirements are scattered across Chapter III Section 2 and as well as other parts of the EU AI act. We will highlight them in the following sections. 

The obligations for accuracy are often presented in tandem with the requirements of \textit{robustness} (rec. 59;  as well as \textit{cybersecurity} (recs. 66; 74; 122; Arts. 13 (2) (b)(ii); 15 (1)). Although the concept of robustness is also not defined in the EU AI act, it refers to high-risk AI system's resilience with respect to possible errors, faults or inconsistencies that may occur within the system or the environment in which the system operates, in particular due to their interaction with natural persons or other systems (Art. 15(4)); whereas cybersecurity refer to deliberate attempts of unauthorized parties to alter the use, outputs or performance of a high-risk AI system by exploiting system vulnerabilities (Art. 15 (5)).
\paragraph{Report Demands of Accuracy}
Many of the EU AI act's requirements on accuracy serve an informational purpose.  First, Article 11(1) requires the providers to draw up technical documentation, to demonstrate compliance with the Section 2. The purpose of the documentation is to support the national competent authorities and notified bodies in the assessment of provider's compliance with the EU AI act. At minimum, it should provide the elements covered in Annex IV, the requirements of which also cover documentation on accuracy of the AI system. Second, it requires the levels of accuracy and the relevant accuracy metrics of high-risk AI systems to be declared in the deployers' use instructions (Arts. 15(3);(13)(2),(3)(ii), indirectly also 13(3)(v)). Overall, the reporting requirements for on accuracy serve the EU AI act's goal of enhancing transparency and fairness of sales of high-risk AI systems (Arts. 13(2)(b)(iii); 15 (3); rec. 74). They are discussed in more detail in Section L.4.

        
    
\paragraph{Accuracy is part of ``performance''}
Both Art. 13 (2)(ii) and Art. 15 (2) list accuracy as a quality of performance of a high-risk AI system. The EU AI act defines performance of an AI system as the ability of an AI system to achieve its intended purpose (Art. 3(18)) see \cite{marin2026your}. In the earlier commentary of the EU AI act, Hacker has proposed replacing the concept of "accuracy" more generally with the term "performance" \cite{hacker2023european}. In the same direction, Engelfriet does not discuss accuracy as a specific criterion, but rather an element of robustness and performance \cite{AIAICTRecht}. Also Nolte et al. pay attention to interrelationship and trade-off between accuracy, robustness and cybersecurity\cite{nolte2025robustness}. However, Vaciago and Constantino treat accuracy as a distinct compliance requirement\cite{AIA15Hart, AIA15Wolterskluwer}.

Confusingly, the EU AI act's compliance requirements also use the terms "accuracy" and "performance" interchangeably. On the one hand, the deployers are required to provide detailed technical documentation about the \textit{degrees of accuracy} for specific persons or groups of persons on which the system is intended to be used (Art. 11 (1); Annex IV (3)). On the other hand, they are obliged to inform deployers about \textit{performance} regarding specific persons or groups (Art. 13 (3)(b)(v)). To add to confusion, Recital 59 suggests that performance and accuracy are different concepts. It says, ``if the AI system is not trained with high-quality data, does not meet adequate requirements in terms of its performance, its accuracy or robustness, or...''. Despite these inconsistencies, we view, in the light of the wording of Art. 13(2)(ii) and Art. 15 (2) and (3) accuracy as a distinct criterion, that is part of the performance of an AI system. 
  We will discuss the meaning of intended purpose in relation to the requirement of accuracy below in Section L.3. 

\subsection{L-accuracy \textcolor{teal}{certifies} the \textcolor{purple}{fitness-to-purpose} of machine learning \textcolor{violet}{systems}}
\label{L:DOES}


\paragraph{Ex Ante, Metaregulatory Nature of EU AI act}

The requirement of accuracy is part of the compliance criteria that the provider of an AI system must comply with before placing a system on the market or putting them into service (Arts. 1(a); 16). In practice, many of the compliance criteria must be taken into account as early as the design phase of an AI system. For example, high risk AI systems must be \textit{designed and developed} in such a way that they achieve an appropriate level of accuracy (Art. 15 (1)). This reflect the ex ante nature \cite{bygrave2022security} of the EU AI act's compliance criteria. 

Before to placing a system on the market or putting it into service, the provider must undergo a conformity assessment procedure (Arts. 16 (f); 43 (1)). The purpose of the procedure is to demonstrate compliance with the requirements of Section 2, to which the requirement of accuracy also belongs to.  The providers are expected to rely on harmonized standards to comply with Section 2 requirements of the EU AI act (Art. 40). The compliance process also involves drawing up the declaration of conformity (Art. 47) and and affixation of a CE marking (Art. 48).

The responsibility for the compliance with the EU AI act vests with the system provider \cite{yeung2022demystifying}. This reflects the meta-regulatory approach of the EU AI act. The specification of the EU AI act's often general compliance rules is allocated to European standard setting organizations. Their capacity to set standards for risks to fundamental rights has been subject to sceptisism\cite{cantero2024artificial, almada2025eu}. Moreover, although non-compliance is subject to fines, metaregulatory approach adopted in the EU AI act puts considerable trust on providers following through the conformity assessment procedure Cf. \cite{coglianese_mendelson_2010}. For high-risk systems qualified by Art. 6(2) and Annex III (2)-(8), the conformity assessment procedure is conducted internally (Arts. 43(1)(a), (2) Annex VI), for high-risk AI systems covered by European product harmonization legislation (under Art. 6(1) and Annex I Section A) the conformity assessment process involves a notified body (Arts. 43(1)(b), (3); Annex VII). 

\paragraph{Accuracy is to be understood with respect to intended purpose}
Harmonized standards, that would specify the compliance with the requirement of an appropriate level of accuracy, have not yet been released. When the law leaves a concept undefined, legal interpretation begins with the ordinary meaning of the word. Following the Oxford English Dictionary ``accuracy'' is ``[t]he state or quality of being accurate'' respectively ``[t]he closeness of a measurement, calculation, or specification to the correct value'' \citep{OED_accuracy_2023}, while ``accurate'' is described as ``conforming exactly with the truth or with a given standard'' \citep{OED_accurate_2023}. Hence, ``accuracy'' is commonly considered to be a truth-related condition. 
    
 However, the use of ``accuracy'' in the EU AI Act suggests a purpose-oriented interpretation of the term. First, the requirement for an appropriate level of accuracy ( Art. 15 (1)) must be read in conjunction with Art. 8, which states that high-risk AI systems should comply with the requirements of Section 2, taking into account their intended purpose. Moreover, the earlier draft of the Art. 15 of the EU AI Act explicitly mentions that ``[h]igh-risk AI systems shall be designed and developed in such a way that they achieve, in the light of their intended purpose, an appropriate level of accuracy'' \citep{europeancommission2021aiact}. Similarly, the annex on a standardization request for the EU AI Act \citep{EC_AI_Standardisation_2024} states that `` 'accuracy' shall be understood as referring to the capability of the AI system to perform the task for which it has been designed. This should not be confused with ‘statistical accuracy’, which has a more limited meaning and is one of several possible metrics for evaluating the performance of AI systems''. Upon commenting on Art. 15, Vaciago proposed an own, also purpose-bound definition of accuracy, defining it as "the AI system's ability to generate correct, reliable and consistent outputs in relation to the specific purposes for which it has been designed and developed."\cite{AIA15Hart}. Similarly, also Finck ties the concept of accuracy to the correctness of a high-risk AI systems' outputs \cite{finck2026eu}.

 The ``intended purpose'' refers to ``the use for which an AI system is intended by the provider, including the specific context and conditions of use, as specified in the information supplied by the provider in the instructions for use, promotional or sales materials and statements, as well as in the technical documentation'' (Art. 3 (12)). It is determined by the AI system provider. The intended purpose of an AI system determines whether or not the AI system is high-risk and subject to EU AI act's compliance obligations. 
 Although the purpose is defined by the provider, the usage context thus cannot be ignored. The reference to instructions, promotional and sales materials, as well as technical documentation highlights their relevance in communicating the intended purpose to the deployers' of the high-risk AI system (Art. 3(12). The provider may use heterogeneous metrics to verify performance \cite{nolte2025robustness} and justify the metrics chosen (Art. 3(18), (32); Annex IV (4))\cite{AIAICTRecht}.
 
 This legal interpretations hence are in line with the suggestion of \citep{four2026derr} to move to felicity conditions in machine learning. ``Felicity'' is a term borrowed from speech-act theory based on the seminal work of Austin \citep{austin1962how}. Instead of whether the outputs of an AI system are (close to) true, felicity conditions ask whether the outputs of the AI system meet their intended purpose. Accuracy is a cornerstone in measuring how well an AI system meets its intended purposes.


\paragraph{Intended purpose is not the ``objective''}
Following the EU AI Act, AI systems generate outputs. The outputs are results of inference, that reflects ``explicit or implicit objectives'' Arts. 3(1)) What is the relationship between objectives of an AI system and the intended purpose of an AI system?

Commission Guidelines explain that ``\emph{explicit objectives} refer to clearly stated goals that are directly encoded by the developer of the system. For example, they may be specified as the optimization of some cost function, a probability, or a cumulative reward. \emph{Implicit objectives} refer to goals that are not explicitly stated but may be deduced from the behaviour underlying the assumptions of he system. These may arise form the training data or the interaction of the AI system with its environment'' \cite{EuropeanCommission2025AIGuidelines}.  

 EU AI act's Recital 12 explains that ``[t]he objectives of the AI system may be different from the intended purpose of the AI system in a specific context''. The guidelines further explain that the AI system objectives are inherent to the internal functioning of the systems ``referring to the goals and tasks to be performed and their results. For instance, a corporate virtual AI assistant system may have objectives to answer questions on a set of documents with high accuracy in (sic) and low rate of failures'' \cite{EuropeanCommission2025AIGuidelines}. In contrast, intended purpose is ``externally oriented and includes the context in which the system is designed and deployed and how it must be operated,'' which includes not only ``the system's internal operation to achieve its objectives, but also through other factors, such as the integration of the system into a broader customer service workflow, the data used by the system, or the instructions of use'' \cite{EuropeanCommission2025AIGuidelines}. In the context of the EU AI act, the concept of the "intended purpose" is thus broader than that of the "objectives" of an AI system\cite{AIAICTRecht}.

While the Commission Guidelines suggest that high accuracy may be an objective of the internal functioning of the AI system, the rest of the systematics of the act do not support the view of where the fulfillment requirement for accuracy (Art. 15(1) is interpretively independent from the concept of intended purpose (Art. 8). As a consequence, the appropriateness of the level of accuracy of an high-risk AI system is also qualified by the use context of an AI system, see also \cite{marin2026your}. 

\paragraph{Intended purpose can be interventional}
 According to the EU AI act, ``AI systems generate outputs, such as predictions, content, recommendations, or decisions  that can influence physical or virtual environments'' (Arts. 3(1)). It is the system's the capacity to influence that triggers concerns about it generating risks of harms. 

 This is evident also from the examination of the categories of high-risk AI systems under the Annex II. For example, a system which is used to decide which students receive special support in school would qualify as high risk AI system in the context of education (\cf \citep{perdomo2025difficult}). These include systems ``intended to be used for the purpose of assessing the appropriate level of education that an individual will receive or will be able to access, in the context of or within educational and vocational training institutions at all levels'' (Annex III, (3)(c).) A system which decides which unemployed persons are assigned to which educational program (\cf \citep{zezulka2024fair}), would also be likely to qualify as a high-risk AI system the context of access to and enjoyment of essential public services and benefits (Annex III (5)(a). 

 In fact, AI systems can as well be framed as not to intervene directly through algorithmic decision-making, but make social predictions which lead to interventions: predict school performance (Annex III (3)(c)) or predict length of unemployment (Annex III (4)(b) or predict a natural persons risk of burglary (Art. 5(d); Annex III (6)(d)).
 
In the light of the definitions of an AI system (Art. 3(1)) and the intended purpose (Art. 3(2)) and categories of high-risk systems (Annex III), we can induct that AI systems can have the purpose to intervene. Whereas this qualification may not hold obvious relevance to the legal community, it is key for understanding how L-accuracy differs from E-accuracy.

\subsection{L-accuracy is inherently \textcolor{teal}{normative} and used for \textcolor{purple}{risk mitigation}}
\label{L:ENDS AND MEANS}

\paragraph{Goals of accuracy in EU AI Act}
When a technical term, such as accuracy, is integrated into legislation it becomes subject to legal interpretations that depart from its original meanings in the technical community \cite{koivisto2024user}. Even if the concept did not have an explicit normative meaning in the field of the machine learning, it now gains one in the field of law. The meaning derives from interpreting the concept in the light of the overall systematics and the objectives of the EU AI act \cite{mollers2022legal}.

First, the reference to an ``appropriate level'' subjects the technical criterion of accuracy to normative evaluation \cite{finck2026eu}.
Second, although the requirement of accuracy is technical, it serves the purpose of AI safety(rec 59;79) \cite{AIA15Wolterskluwer}. In the broader context of Art. 15, which Vaciago qualifies Accuracy as a "safequard condition", that contributes to the goal of mitigating risks stemming from technical, computational and cybersecurity dimensions \cite{AIA15Hart}. 
Third, in conjunction with the requirements of robustness and cybersecurity, accuracy serves the goal of supporting the uptake of trustworthy AI (\cite{finck2026eu, AIA15Wolterskluwer}, Art. 1(1)). 
Fourth, as mentioned above, accuracy connects to the objectives for transparency and fairness of sales of high-risk AI systems (Arts. 13(2)(b)(iii); 15 (3) rec. 74), which is reflected in the Act's obligations to disclose the accuracy metrics deployed (Arts. 11(1); Annex IV (2)(g) and(3); 13(2)(b)(iii); 15 (3)), see also (\cite{AIA13Hart,AIA15Hart}. 
Fifth, as discussed above, the appropriate level of accuracy is qualified by the intended purpose of the system (8(1) Art. 15(1). It is part of system's performance - ability to reach achieve the intended purpose (Art. 3(18). These qualifiers give the normative concept of L-accuracy a particularly teleological nature.
Sixth, according to Art. 8 (1), \emph{the risk management system} required by Art. 9 should be taken into account when ensuring the compliance criteria of Chapter III Section 2, including accuracy.
Finally, like other requirements for high-risk AI systems, the demand for accuracy is deemed necessary to mitigate the risks for health, safety and fundamental rights (rec. 66) \cite{AIA15Wolterskluwer} For example, the recitals highlight the importance of accuracy in mitigating fundamental rights risks posed by using AI in law enforcement. Together with reliability and transparency of AI systems, AI system's accuracy helps to avoid its adverse impacts, to retain public trust and to ensure accountability and effective redress (rec. 59).  
 


\paragraph{Accuracy ``throughout the lifecycle''}

Besides the explicitly stated goals, also the wording and systematics of the EU AI act also shape the normative meaning of the accuracy requirement. The EU AI act mandates the providers of high-risk AI systems to consider accuracy already in its design and development stage (Art. 15 (1)) and to ensure that the system performs with  a consistent level accuracy of throughout the lifecycle (Art. 15 (1); rec 73) see also \cite{nolte2025robustness}. This corresponds to overall life-cycle based risk-management approach of the EU AI act. Also the obligations for risk-management systems (Art. 9 (2); rec. 65) and human oversight (rec. 73) are meant to cover the system lifesycle. In the recitals, the logic extends to compliance requirements for GPAI Models with systemic risks (recs. 114-115).

Among the compliance criteria for high-risk AI systems, the obligation of the risk management system (Art. 9), holds also higher-order, teleological interpretive meaning. According to the Art. 8, the risk management system shall be taken into account when ensuring compliance with other requirements. Being a key tool for identifying and migitating a high-risk system's risks of health, safety, and fundamental rights throughout the lifecycle the system (Art. 9), it also shapes the interpretation of the requirement for accuracy. Most importantly, the EU AI act established an obligation for continuous identification of the known and the reasonable foreseeable risks that the system can pose on health, safety and fundamental rights and adoption of risk management measures to mitigate them (Art. 9 (2)(a)(b)) together with a requirement to establish a post-market performance monitoring system (Arts. 9(2)(c); 17 (h); 72). On this basis, the requirement of accuracy and the broader notion of \emph{performance} should be interpreted in the light of the overarching goal of mitigating risks throughout the life-cycle of an AI system. 

Generally, providers are required to consider the \textit{interaction between the individual compliance criteria} of Chapter III Section 2. Consequently, it should be considered that the criterion of accuracy "extends not only to the algorithms and computational processes underpinning the system but also the quality of input data, the design logic, and the operational conditions in which the system is developed." \cite{AIA15Hart}.  For example, the risk management measures shall give due consideration to the effects and possible interactions from the combined application of compliance requirements (Art. 9 (4), Chapter 3) and describe the decisions about any possible trade-off made regarding the technical solutions adopted to comply with the requirements set out in Chapter III (Annex IV (2)(b)). For a review of how Art. 15 interacts with other requirements of Chapter III, see \cite{AIA15Hart}.

The compliance requirements for accuracy and performance in Chapter III Section II can be divided into three categories based on the life-cycle stage they focus on: 1) consolidating the expected level of performance through validation and testing, 2) managing expectations by communicating the expected level of performance, 3) enabling the monitoring of expected performance. 
    
    \begin{enumerate}
        \item Prior to bringing the AI system to the market, performance is established through testing and validation procedures and appropriate metrics to ensure their consistent functioning (Arts. 13(3)(b)(ii); Annex IV (2) (g); 9 (6)). Further compliance conditions on data governance and the quality of training and validation and training data are included in Art. 10 of the EU AI act\cite{AIA15Hart, AIA13Hart, AIAICTRecht}.
        \item The expected level of performance is communicated through technical documentation and transparency measures: 

        \begin{itemize}
    \item Deployer's instructions should be intelligible \cite{AIA13Hart} and state the characteristics, capabilities, and limitations of performance of the high-risk AI system (Art. 13 (3)(b)). These encompass a declaration on the levels of accuracy and the relevant accuracy metrics (Art  15(3)),against which the high-risk AI system has been tested and validated(Art. 13 (3)(b)(ii)).
    \item Transparency towards deployers about any known or foreseeable circumstances that may have an impact on that expected level of accuracy (Art. 13(3)(b)(ii)).
    \item Providing a high-risk system to the deployer in a manner that allows natural persons to whom human oversight is assigned to detect and addressing anomalies, dysfunctions and unexpected performance (Art. 14(4)(a)).
    \item Description [...] of the output quality of the system (Art. 11 (1); Annex IV (2)(b)).
    \item Detailed information about the overall expected level of accuracy in relation to its intended purpose (Art. 11(2); Annex IV (3)).

\end{itemize} 
        \item The providers must keep the technical documentation at disposal for 10 years after placing an AI system on the market or putting into service (Art. 18(1)(a)). The documentation  provides the foundation for post-market monitoring and evaluation of the system. The monitoring system in-itself must also be documented by providing a detailed information about the monitoring, functioning and control of the AI system, in particular with regard to its capabilities and limitations in performance (Annex IV (3)) and of the system in place to evaluate the AI system performance in the post-market phase” (Annex IV 9). The obligation to keep logs supports the monitoring of the fulfilment of the accuracy criterion (Arts. 12 and 19) \cite{AIA13Hart,AIA15Hart}. Moreover, the deployer should be enabled to properly understand relevant capacities and limitations of the system with respect to its performance (Art. 14 (4)) to perform human oversight.
    \end{enumerate}

\paragraph{Validity Arguments for accuracy are central}
When read in the light of the goals and structure of the EU AI act, the criterion for accuracy must be  
 \begin{itemize}
     \item interpreted in the light of several normative concepts of the act
     \item understood as a criterion, the fulfillment of which extends throughout different phases of the AI system's lifecycle
 \end{itemize}
This approach to accuracy is not intuitive to AI developers who understand accuracy as an outcome or a metric. Instead, EU AI act focuses on the process of developing a compliant AI system. It does not focus on the outcome. Instead, warrants for the validity are given by a following a certain process (\eg, good training data used) and not the measurement itself. As a consequence, we argue that the  EU AI act requires the providers to give \emph{justifications for the validity of accuracy}. The provider is required to report or fulfill accuracy needs not only to state the choices made, but argue why those choices are legitimate and well-founded.
We exemplarily show in the following how this is made explicit within the EU AI act for \emph{accuracy metrics} and \emph{testing data}.

The EU AI act articulates an obligation for the transparency of the used accuracy metric: ``The levels of accuracy and the relevant accuracy metrics of high-risk AI systems shall be declared in the accompanying instructions of use'' (Art. 15 (3)). Similarly, Art. 13(3)(b)(ii) requires transparency about the level of accuracy and its metrics against which the system has been tested and validated towards the deployers.

Beyond mere reporting, the EU AI act assumes a plurality of accuracy metrics that should be determined \emph{based on the intended purpose} of the high-risk AI system and the state of the art (Art. 8(1); Art 15 (2); Annex IV (4)). In particular, technical documentation should include a description of the appropriateness of the performance metrics (Annex IV (4)), to which accuracy belongs to (\ref{L:IS}).

Besides obliging the providers to document and justify the accuracy metrics,
EU AI act specifies that accuracy is established through the \emph{process of validating and testing} the AI system. It requires the provider to provide a detailed description of the validation and testing procedures used, including [...] data used and their main characteristics; metrics used to measure accuracy, robustness (Annex IV (2) (g)).
More specifically, Annex IV (2)(g) mandates them to disclose ``the validation and testing procedures used, including information about the validation and testing data used and their main characteristics''. 
In addition, the providers must disclose ``relevant information'' about the testing data to the deployers (Art. 13(3)(b)(iv)), where ``testing data'' refers to data used for providing an independent evaluation of the AI system in order to confirm the expected performance of that system before its placing on the market or putting into service (Art. 3(32)).

Again, disclosure of the testing procedures and data is not sufficient. EU AI act Article 10 requires that training, validation and testing data sets to be relevant, sufficiently representative, and to the best extent possible, free of errors and complete in view of the intended purpose. They shall have the appropriate statistical properties, including, where applicable, as regards the persons or groups of persons in relation to whom the high-risk AI system is intended to be used. Those characteristics of the data sets may be met at the level of individual data sets or at the level of a combination thereof (Art. 10(3)).

In short, providers must provide normative reasoning for their design choices how they evaluate the accuracy of an AI system. In fact, the commission is mandated to ``encourage, [...] the development of benchmarks and measurement methodologies'' (Art. 15(2)), hence to offer technical guidelines and support for the providers to follow their obligation.


\subsection{L-accuracy \textcolor{teal}{exists to protect} subgroups and individuals}
\label{L:PARTS AND WHOLES}


EU AI act qualifies accuracy with with respect to individuals and groups.
First, deployers' instructions require provide, where appropriate, information about its \emph{performance} regarding \emph{specific persons or groups of persons} on which the system is intended to be used (Article 13(3)(b)(v)). Second, the technical documentation on the AI system must include ``detailed information about the monitoring, functioning and control of the AI system, in particular with regard to: its capabilities and limitations in performance, including \emph{the degrees of accuracy for specific persons or groups of persons} on which the system is intended to be used ( Annex IV(3), emphasis added).

The calls for providing accuracy with respect to individuals may stem from the legal system, generally, being structured around rights of individuals. For example, EU fundamental rights apply to individuals residing in the EU. However, it is unclear why the provisions refer to specific persons -- especially considering that providers must comply with the provision before the AI system has been brought to market or put into service.

AI systems can have scalable effects on groups of people, which may be unobserved by the individuals affected by them. Already EU AI act's impact assessment recognized the tension between the individual and AI: ``[...] AI is often used to sort and classify traits and characteristics that emanate from datasets that are not based on the individual concerned'' \cite{Impactassessment}. Recognizing this, the EU AI act is sprinkled with considerations for groups of people with regards which the system is intended to be used (Art. 4; Article 13(3)(b)(v); Annex IV (2)(b)) including the above-mentioned obligations. In addition, the Act mentions harms to groups (Art. 7 (2)(f);(j) vulnerable groups (Arts. 5 (b), (c); 9 (9), 60 )(g); Art 79 (2), 98 (2) (e)) especially with respect to biased AI systems (rec. 67); deployers' obligations to consider fundamental rights impacts of groups (Art. 27).

Providing information about accuracy with respect to specific individuals and subgroups, ex ante, before the system has been brought to the market may be impossible. Moreover, as discussed in Section E.5, it may be technically infeasible. Whereas compliance with Art. 13 (3)(v) is conditional on "appropriateness", which may include considerations of feasibility, the obligation for technical documentation is not subject to such qualifiers (Art. 11(1); Annex IV (3)).

 

\paragraph{Summary}
One can summarize that accuracy is one of the key qualities of high-risk system's performance. Accuracy is a measure of a high-risk AI systems capacity to achieve it's intended purpose. Moreover, it is a system quality that contributes to the fulfilment of the normative goal of mitigating risks to health safety and fundamental risks where ensuring consistent performance over the lifetime of the system is instrumental to the fulfilment of that goal. Accuracy is established through testing and validation of a high-risk AI system, a process which should be informed by risks to health, safety and fundamental rights of that system. The metrics to establish accuracy are heterogeneous and context-dependent, yet should be reflective of the state of the art. Transparency about the metrics ensures the verifiability of system's compliance and establishes a baseline for system's expected, intended performance as well for identifying deviations from that performance by the deployers as well as through post-market monitoring.

\renewcommand{\thesection}{E}
\section{What is E-Accuracy? ML Engineer's View on Accuracy}
\label{E:What is E-accuracy? -- Appendix}

Accuracy is so central and standard to machine learning that the question what it \textit{is} feels inappropriate. Nevertheless, carving out the facets of E-accuracy will serve several purposes here.
We make explicit what is unarticulated, implicit knowledge in machine learning. In doing so, we emphasize aspects which might seem marginal to the machine learning community, which, however, turn out to be fundamental anchors for tensions with respect to L-accuracy. Hence, we try to sketch E-accuracy for a legal community. In particular, we argue that:
\begin{enumerate}
    \item Accuracy is a formal, empirical concept.
    \item Accuracy defines performance of ML models or systems.
    \item Accuracy is a comparative measure of performance.
    \item Accuracy is a statistical measure, particularly appropriate for statistical systems such as created by machine learning.
\end{enumerate}

\renewcommand{\thesubsection}{E.\arabic{subsection}}
\subsection{Introduction to \textcolor{teal}{engineering} of (computational and statistical) \textcolor{purple}{technologies}}
\label{E:CONTEXT}
``AI'' is a buzzword currently filling media. Large parts of the ``hype'' \citep{widder2024watching} are based on advancements in the field of machine learning. What is machine learning? And, how is machine learning done? Below, we present our own somewhat colored account, which is by no means exhaustive.

\paragraph{Machine Learning Creates Statistical Technology}
Imagine, you have to the task to create a system which gets images and needs to identify whether the image shows a car or not. Naively, one might come up with a set of rules like ``If an image shows an object with tires, it shows a car.''\footnote{This mapping rule is highly-stylized. Clearly, it would require to first define how to detect tires.} to create this system. The central (and only?) twist to the machine learning approach is the move from hard-coded mappings to data collection which inform the choice of mapping. Instead of a mapping based on rules, now the mapping is obtained by ``If an image looks similar to some of those car images, it shows a car.'' That is, machine learning creates statistical systems, much in the spirit of the ``statistical method''. ``In the [...] statistical method the [...] conclusions rest solely on empirically established relations between data and the condition or event of interest.'' \citep{dawes1989clinical}\footnote{What we call ``statistical'' is sometimes called ``actuarial'' in literature. The reason is the historical interconnection between statistics and insurance \citep{gigerenzer1990empire, desrosieres1998politics, lehtonen2014insurance}.}

\paragraph{Statistical Technology Needs Reference to Aggregates}
Usually, during \emph{training} a mapping is adapted to fit a set of training instances, \eg, \citep{bishop2006pattern, shalev2014understanding, goodfellow2016deep}. That is, among a set of potential mappings, which are most often called \emph{models} in machine learning, one which performed best is chosen. By ``best performance'' we mean, that the model achieved a high \emph{accuracy} (see the next sections), on a set of instances for which we knew the desired outputs. Exemplarily, among a set of mappings from images to ``car/no car'' the mapping which most correctly guessed the class on a set of images for which the classes are known, is chosen. Hence, the model obtained by training has undergone statistical selection. The outputs generated by the mapping, \ie, the model, can only be rationalized by reference to the aggregate, the defining feature of statistical technologies \citep{four2026derr}.

\paragraph{Machine Learning is Engineering Computational and Statistical Technologies}
The simple scheme of training is extended and refined. Critically, the mechanisms are implemented on computing machines. The result is a technology capable of generating predictions for medical purposes \citep{esteva2017dermatologist}, recommendations for social media \citep{covington2016deep}, or images for video generation \citep{ho2020denoising}.
Historically machine learning as had an ambiguous status as 
a technical discipline. Is it a science? (clearly not --- it offers no new scientific truths about the world). 
Is it just mathematics and programming? (No, because 
while it uses those tools, it ingests data and acts upon the world). 
Is it a technology? Yes, although a rather immature one, analogous 
to the aeronautical technology of a century ago 
\citep{vincenti1990engineers}  --- it is still evolving its
engineering practices and standards.

\subsection{E-accuracy is a \textcolor{teal}{formal and empirical} measure}
\label{E:IS}
Classically, in machine learning accuracy is (most often) defined as the mismatch rate of a binary predictor with a class label, \eg, \citep[Section 2.2]{shalev2014understanding}. In other words, accuracy is one minus the relative number of incorrect classifications of objects to belong to the class $0$ (respectively $1$), when they belong to the class $1$ (respectively $0$). We refer to this concept as \emph{classical accuracy} in the following.

More precisely, classical accuracy is comprised of three ingredients: binary outputs of a machine learning model $\hat{Y}_i \in \{ 0,1\}$ , desired binary outputs $Y_i \in \{ 0,1\}$, and the comparison function which calculates, \ie,\footnote{The Iverson-bracket which evaluates to $1$ if the bracketed term is true and $0$ otherwise, is denoted $\llbracket \cdot \rrbracket$.}
\begin{align}
\label{eq:classical accuracy}
    \operatorname{ClassyAcc} = \frac{1}{|I| } \sum_{i \in I} \llbracket \hat{Y}_i = Y_i \rrbracket.
\end{align}
Note that we use $i \in I$ as an indicator over a finite set of instances $I$. Hence, interestingly, the most immediately provoked answer to the question ``what is accuracy'' is answered by responding to ``how do we measure accuracy''.

Machine learning systems obviously do not only produce binary outputs. The concept of accuracy developed accordingly. Let $\hat{Y}_i$ be some output of a machine learning model, \eg, a rank of an instance or a real-valued score, and $Y_i$ its desired counterpart. A more general, but still mathematical form, of accuracy could read as,
\begin{align}
\label{eq:average accuracy}
    \operatorname{AvgAcc} = \frac{1}{|I|} \sum_{i \in I} u(\hat{Y}_i, Y_i),
\end{align}
where $u$ is a function, which expresses how well the generated and desired output match. Machine learning scholars actually prefer to talk about a loss function $\ell$ which expresses how well the generated and desired output differ. That amounts to a simple sign-flip. The higher the loss, the worse. The higher the accuracy, the better.

In the era of (large) language models (LLMs) the outputs can take even more semantically loaded forms such as text. In this case, the desired outputs sometimes only refer to properties of the output, but not the actual output itself. For instance, it does not matter whether the machine learning model produces the letter ``p'', but rather whether it mentions a word which is semantically related to ``penguin''. Hence, ``desired output'' should be understood broadly, as well as potentially ambiguous in certain situations. For instance, whether the desired output is ``penguin'' or ``spheniscidae'', the latin family name of penguins, is actually not clear.

Nevertheless, there are certain features shared by (most) definitions of accuracy: (a) They are \emph{output-focused} and neglect the internal workings how the output was generated. (b) They are \emph{statistical}, in the sense they aggregate over a set of instances. This does not need to be the average as for the two mathematical forms given (\eqref{eq:classical accuracy} and \eqref{eq:average accuracy}). (c) They require a formal comparison metric and a data set of inputs to the machine learning model in addition to the desired outputs. We'll elaborate on those two ingredients in the following.

\paragraph{Accuracy Metrics Abstractly Capture Utility}
Even for binary outputs, the comparison of produced binary outputs with desired binary outputs can take a variety of forms, \eg, true positive rate, false positive rate, specificity, $F_1$-score, which are based upon counting the amount of instances for all possible four constellations of produced and desired outputs summarized in the confusion matrix \citep{dobson2024confusion}.
For instance, the true positive rate is the ratio of the amount of instance with $\hat{Y}_i = 1$ in comparison to the amount of instances $Y_i = 1$. In other words, the true positive rate is the ratio of correctly as positive predicted instances with respect to all positive instances. How to choose among the those metrics?

Essentially, different scenarios involving distinct utilities of involved parties motivate different choices.
Imagine an X-ray image classifier for pneumonia.
If the purpose of the system is to sort out clearly uncritical patients, then instead of classical accuracy the true positive rate could be more appropriate to measure the ``accuracy''. For a high true positive rate, only clearly uncritical patients are classified as such. While the patients might be satisfied with this choice, the hospitals might ask for efficiency gains and demand for high true negative rate, which might lead to less of the costly post-classification diagnosis by doctors.
The accuracy metric chosen is often somewhat aligned to the use of the model's outputs (\cf Section~\ref{E:DOES}).

Historically, the link between the use of a statistical system and its evaluation has at least already been established by Abraham Wald in 1950 \citep{wald1950statistical}. That accuracy metrics capture abstract utilities is hence a development along this trajectory, which obviously generalizes beyond binary outputs \citep{gneiting2007strictly, gneiting2011making, four2026derr, derr2025three}.

\paragraph{Data as ``Undervalued Given Truth''}
Accuracy is not only the metric chosen. Accuracy requires test data. More precisely, accuracy requires outputs of a machine learning model based on inputs and desired outputs. The produced and desired outputs are then compared. In that sense, accuracy is empirical.

Despite its centrality, machine learning scholars tend to treat the (test) data as a ``truth'' \citep{aroyo2015truth, moss2021objective} ``given'' \citep{williamson2024rhetoric} to them, while being ``undervalued'' \citep{sambasivan2021everyone}.

Large parts of currently used test data, usually packed into \emph{benchmarks} (Section~\ref{E:ENDS AND MEANS}), are created by ``convenience samples'' from the internet, or artificially generated, and then having (non-)experts, or other AI systems, annotate the retrieved images, texts, videos and others with respect to a relatively abstract task \citep{deng2009imagenet,lin2014microsoft, rajpurkar2016squad, wang2018glue}. For instance, the famous ImageNet dataset has been collected from various sources by querying image search engines. Those images were then categorized based on terms in WordNet \citep{fellbaum1998wordnet} using Amazon mechanical turk\footnote{Amazon mechanical turk is ``a crowdsourcing marketplace that makes it easier for individuals and businesses to outsource their processes and jobs to a distributed workforce who can perform these tasks virtually.'' (\url{https://www.mturk.com/}, 24th of March 2026)}.

``Truth'' of the data is presumably established by using ``raw data'' \citep{gitelman2013raw} and human annotations \citep{aroyo2015truth}.
That data set choice can be a choice between world representation shaped by, \eg, limited human capabilities \citep{hernandez2020ai} or normative tendencies \citep{pine2015politics, dotan2020value}, plays a minor role. Instead, scholars often claim to work with ``real-world'' data \citep{birhane2022values}.
Hence, machine learning feeds its ``truth fetish'' \citep{moss2021objective, four2026derr} with ``true labels'' and ``ground truth''.

Furthermore, the processes, journeys and stories of creating data \citep{leonelli2020data} are often cut short. The ImageNet data set consists of images which were produced for a variety of purposes, but not necessarily for evaluating computer vision models \citep{deng2009imagenet}. The exact creation process of MNIST, a data base of handwritten digits \citep{deng2012mnist}, has been lost \citep{yadav2019cold}. Even in the literature on \emph{data selection}, which by its definition treats data as a design choice, we find traces of ``data as a given''. For instance, the authors in \citep{albalak2024a} write as a first sentence, ``Data selection is a long-standing challenge of machine learning where, \emph{given a collection of raw data}, the goal is to design a dataset that is optimal under some objective function [...]'' (emphasis added).

Finally, despite its prominence, data work is undervalued. Data workers are paid significantly less in comparison to machine learning engineers, and often exploited in precarious labour markets \citep{gray2019ghost, hao2025empire}.
The methodological importance of data selections is neglected in some high-stakes deployment scenarios \citep{sambasivan2021everyone}.

In conclusion, test data and accuracy metric are the elementary ingredients for E-accuracy. E-accuracy is both formal (through ``useful'' mathematically defined metrics) and empirical (through ``given'' data).


\subsection{E-accuracy \textcolor{teal}{defines} the \textcolor{purple}{performance} of machine learning \textcolor{violet}{models}}
\label{E:DOES}
E-accuracy is used in machine learning. \citet{birhane2022values} claim it is even \emph{a} central value of machine learning. But {what is E-accuracy used for}?  What does E-accuracy do?

\paragraph{Accuracy as an Abstract Performance Objective}
In 54 of the 77 oral papers at the main conference track of Neural Information Processing 2025\footnote{This number has been acquired by using a script generated with the help of Claude Sonnet 4.6 by Anthropic. \emph{Orals} are highlighted papers. Neural Information Processing Systems (NeurIPS) is one of the largest and renowned machine learning conference and publication venue.}
the word ``accuracy'' is mentioned. Most often, ``accuracy'' is used within experimental sections in which a suggested machine learning method is compared to the ``state of the art (SOTA)''.\footnote{In $21$ of the $77$ oral papers ``state of the art'' or ``SOTA'' is mentioned.}
Accuracy is used to demonstrate the ``goodness'' of a method. As the problems to be solved are rather abstract tasks, such as image classification, image segmentation, explaining, question-answering, or reasoning, accuracy defines a rather abstract and general notion of performance. The exact deployment scenario of a method is often kept aside. In particular, ``accuracy'' is taken as an objective on which machine learning models are optimized. That is reflected in numerous leaderboards, benchmarks and arenas, \eg, on HuggingFace\footnote{ \url{https://huggingface.co/docs/leaderboards/index}}, CodeSOTA\footnote{\url{https://www.codesota.com/}} or AI Arena\footnote{\url{https://arena.ai/leaderboard}}. Those competitions serve the comparison of machine learning models by heavily resting upon accuracy measurements. Additionally, that is reflected in ``competitions'' tracks at conferences\footnote{For instance, the competitions track at NeurIPS 2025, \url{https://nips.cc/virtual/2025/loc/san-diego/events/Competition}} which explicitly ask for new challenges to compare machine learning methods against each other where accuracy is often at the core of the comparison. And finally, it is reflected in workshops such as ``I can't believe its' not better'' at large conferences\footnote{ICBINB, \url{https://icbinb.cc/workshops.html}} which turn against the ``accuracy objective'' and explicitly claim ``to create and foster a community that goes beyond benchmark climbing and new algorithms with bold numbers.''
The use of accuracy as performance objective is so central that \citet{birhane2022values} write,
\begin{quote}
    generic success terms, such as `success', `progress', or `improvement' are used as synonyms for performance and accuracy.
\end{quote}

\paragraph{Accuracy as Decision Support for Which Method to Use}
The closer to deployment, the more accuracy is considered with respect to a more concrete task to be solved. There is a gradual transition from research over development to deployment. For instance, even though the AI Arena\footnote{\url{https://arena.ai/leaderboard}} provides leaderboards for relatively abstract tasks, the rank has a concrete influence on deployment and use decisions \citep{chen2026leaderboard}.\footnote{Admittedly, the leaderboards of AI Arena are not the result of comparing accuracies but rather of ratings of users.} Other benchmarks, such as WeatherBench \citep{rasp2020weatherbench} are by design of its data curated for concrete and actual tasks, even though the validity of model performance on this benchmark for actual deployment can be questioned \citep{freiesleben2025benchmarking}.

More generally, ``being accurate'' does not necessarily translate to successful deployment \citep{paleyes2022challenges}. Accuracy is then rather used as ``a health check, to make sure the algorithm does what we want to'' \citep{bernardi2019150}. Hence, in the context of deployment, accuracy is used to decide which machine learning system to implement, or whether to implement one. But, it is not necessarily an objective.

\paragraph{Accuracy is Limited in Defining Performance}
\label{para:E:accuracy is limited in defining performance}
As stated, model accuracy does not necessarily translate into successful application \citep{burgess2017AI, bernardi2019150, vanian2020How, raji2022fallacy, kapoor2023leakage, wang2024against}. The reasons for this failure of transferability are manifold:
\footnote{The dimensions spelled out here are heavily inspired by \citep{wang2024against}. Different to their question, ``what are reasons why predictive optimization fails'', we ask, ``what are reasons why accuracy does not transfer to deployment performance''.}
\begin{description}[leftmargin=4pt, labelindent=0pt]
     \item[Accuracy Metric Mismatch] The accuracy metric does not reflect utility in the use-case \citep[Section 4]{marin2026your}. For instance, classical accuracy can be inappropriate when misclassification of a certain class is particularly costly, \eg, incorrectly predicting life-threatening cancer is more severe than incorrectly predicting a cold \citep{elkan_foundations_2001}. 

    \item[Data Set Mismatch] The test data on which accuracy is measured is not ``representative'' for the case at hand. Usually, distributions shifts are claimed to be the cause for this mismatch \citep{koh2021wilds}. Because the distribution of data ``in-the-wild'' is ``shifted'' from the test data distribution, accuracy claims are not transferable. Those shifts can be temporal changes of the data intrinsic to the nature of the problem, \eg, water resources in the times of anthropogenic climate change \citep{milly2008stationarity}, performative effects, \ie, changes of the data due to the deployment of a prediction tool \citep{perdomo2020performative} as in social context such as traffic prediction, or restricted data access during testing \citep{corey2019benchmark, corbett2023measure}.
    
    \item[Target-Construct Mismatch] On a more fundamental level, the output of a machine learning system is often in a vulnerable relation to the purpose of the system. For instance, a machine learning tool with the purpose to assess the risk of a person to become victim of robbery, \ie, an abstract, potentially unobservable \emph{construct}, can only produce outputs referring to concrete, observable \emph{targets} such as police reports, \cf \citep{eaglin2017constructing, bao2021compaslicated}. Often the link between construct and target is weak, \eg, job performance and working hours \citep{campbell1990modeling, robotham1996competences} or economic welfare and gross domestic product \citep{macekura2020mismeasure}. Hence, the problem is that \begin{quote}
        Target-construct mismatch contradicts developers’
    claim of accuracy, because no matter how good a model may be at predicting a target variable, if
    the target variable deviates from the construct of interest,then any claims about the accuracy fall
    short due to measurement error. \citep{wang2024against}
    \end{quote}
    
    \item[Intervention Mismatch] Accuracy is assessed by formally comparing outputs of a machine learning system with desired outputs. Hence, the outputs of the system are treated as non-interventional. They have no effects on the desired outputs. But, machine learning systems can have effects on the data \citep{mendler2024engine}. In particular, the success of the interventional role of machine learning outputs can \emph{not} generally be measured by accuracy, \eg, \citep{bernardi2019150,lemmens2020managing, hutchinson2022evaluation, saxena2023rethinking,wang2024against}. 
    A standard (and debated) example is recidivism risk prediction \citep{chouldechova2017fair, dressel2018accuracy}. If the risk prediction is used to retain incarcerated persons, there is an immediate effect of the output of the machine learning system on the variable it predicts. Interventional measures of success are then, so is argued, more appropriate to ``use data as part of a reflexive practice of testing hypotheses in the service of achieving near and long term goals.'' \citep{barabas2018interventions}. However, accuracy can be supportive to ensure interventional success \citep{fernandez2022causal, fischer2025value, fischer2026empirically} (\cf Section~\ref{clash:a plea for decision accuracy}).

    \item[Broader Understanding of Performance] Even though, we argue that E-accuracy defines ``performance'', ``performance'' can be understood more broadly. Dimensions such as, energy consumption, necessary computing resources, or maintenance costs, are sometimes considered to be part of performance \citep{ethayarajh2020utility}.\footnote{Surprisingly, even literature on energy consumption of machine learning systems define ``performance of AI systems'' only with respect to generated outputs, \eg, \citep{schwartz2020green, henderson2020towards}. They rhetorically undermine their own goal of considering the reduction of energy consumption a central component of ``goodness'' and ``quality''.} In contrast, accuracy is output-focused.
    Essentially, the difference marks the move from machine learning \emph{models} to machine learning \emph{systems}. Systems, in contrast to models, encompass more facets of the technology, \eg, by taking into account the hardware.
    
    

\end{description}
The named limitations of accuracy in measuring performance, except for ``Broader Understanding of Performance'', can be traced back to questions of validity. The measurement of accuracy does not entail the method to be accurate beyond the very particular measurement scenario. \textbf{Missing validity limits the scope of accuracy in defining model performance.}
However, following the example of \citep{freiesleben2025benchmarking}, we need to take into account the purpose and use of accuracy to argue about validity claims. In the following section, we argue that accuracy essentially serves a comparative goal and hence is already useful despite a limited range of validity arguments.

\subsection{E-accuracy is inherently \textcolor{teal}{statistical} and used for \textcolor{purple}{ranking models}}
\label{E:ENDS AND MEANS}
Much of machine learning has become engineering -- the creation and maintenance of technologies of computational, \emph{statistical} predictions, recommendations and generations of artifacts (Section~\ref{E:CONTEXT}). Accuracy is a \emph{statistical} measure (Section~\ref{E:IS}).
Is the \emph{statistical} measure of accuracy only defined for \emph{statistical} systems producing outputs? We now argue that while defined, it is of little use, as statistical systems are favored. 
The thrive for the better statistical method inherent to E-accuracy recurs in the central use of it, \emph{benchmarking}, \ie, ranking models based on E-accuracy.

\paragraph{Statistical Systems Are Favored by Statistical Metrics}
Nothing in the conception of accuracy in machine learning prevents us from measuring the accuracy of non-statistical systems. The output generating system can for instance be a physical simulator, or humans. For instance, in his known studies Paul Meehl, evaluated the accuracy of clinicians in comparison to statistical systems for diagnosis and prediction in particular in psychatric settings \citep{dawes1989clinical, dawes2014statistical}. The critical finding was that statistical systems outperformed clinicians, in terms of higher accuracy. That should be of little surprise though, as statistical systems are exactly the ones which are optimized for achieving high accuracy \citep{recht2025actuary}. Hence, the statistical measure of accuracy is \emph{defined} beyond statistical systems, but by its internal workings, statistical systems are implicitly ``favored''.

\paragraph{Progress is Measured in Benchmarks}
Indeed, statistical systems, as engineered by machine learning, follow countless optimization steps, where ``improvements'' are largely tracked by \emph{benchmarks} \citep{Donoho2024Data, hardt2025emerging}.\footnote{In doing so, machine learning mimics earlier nascent fields of engineering, which started with an obsession for empirical, monolithic performance evaluations, but gradually evolved to a richer and more holistic form of evaluation, \cf \citep{vincenti1990engineers}.}
Benchmarks originated from data sets which were shared by researchers to train and test their machine learning models. For a unified comparison, researchers reused the same data set for training their own method and reporting the obtained accuracies \citep{orr2024ai, hardt2025emerging}.
The potentially most famous example is the ImageNet challenge \citep{deng2009imagenet, russakovsky2015image} which substantially triggered the research on deep learning \citep{krizhevsky2012imagenet, goodfellow2016deep, he2016deep}.
ImageNet comprises around 50 million images labeled by (hierarchical) categories such as ``husky'' or ``sailing vessel''. Machine learning methods compete for achieving correct categorical classifications on a subset of hold-out images, \ie, images which where not shown to the machine learning system during training.

The use and meaning of benchmarks has changed quite substantially over time \citep{hardt2025emerging}. Nevertheless, the comparative nature of benchmarks, \ie, one method is favored over another when it achieved higher accuracies on benchmarks, remains constant. We'll argue that the usefulness of E-accuracy is warranted in this endeavour as it only needs to be comparative. 

\paragraph{Use of Accuracy in Machine Learning (Benchmarking) only Requires Relative Comparison}
E-accuracy is used to select machine learning models, either for choosing the ``better'' model in deployment or for making ``better'' models in development (Section~\ref{E:DOES}). For this use, it if sufficient to rely on relative comparison. That is, E-accuracy only needs to express which machine learning model is better than another.

That largely simplifies the argument for the validity and usefulness of the E-accuracy measurement. The exact number of E-accuracy is not necessary for its use. The ranking of E-accuracies of different methods and models, however, should be valid beyond a specific accuracy measurement. That validity claim is relatively well substantiated, as we argue in the following.

\paragraph{Reduced Validity Claims for Relative Comparison and Competition are backed}
On the one hand, there is strong experimental evidence \citep{recht2019imagenet, yadav2019cold, miller2020effect, miller2021accuracy, salaudeen2024imagenot} that even though accuracy scores obtained by machine learning models on the test data are not indicative beyond the given setup, the rankings of the accuracies are relatively stably preserved. For instance, the ranking which model performs best, second best, etc., on ImageNet is relatively well preserved despite changes of the test set to entirely newly annotated images \citep{recht2019imagenet}.
On the other hand, there is theoretical work that suggests that validity claims for relative comparison allow for otherwise questionable practices such as \emph{test-set reuse} \citep{blum2015ladder, hardt2017climbing}. More concretely, even if machine learning scholars optimize their models based on test set accuracy for ImageNet, exploiting the fact that they can test and re-test again and again, the claim that the ranking of the best models is representative beyond the test data is backed. 

\paragraph{Validity Claims Beyond Comparison are Hard to Hold}
However, mere accuracy scores have proven unreliable to be generalized to unseen test data, \eg, \citep{recht2019imagenet, liao2021we}. \citet{suhr2025stop} state that ``specific accuracy numbers are essentially meaningless as they are easily swayed by small random fluctuations''. 
Most often the \emph{internal validity}, what justifies that the measured accuracy is meaningful within the measurement context, is questioned.
For instance, test-set reuse, \ie, testing a model and then optimizing the model using the knowledge of the test result, can erode standard validity statements of learning theory \citep{dwork2015reusable, blum2015ladder, mania2019model, liao2021we}.
In addition, the ``data set mismatch'' problem (Section~\ref{para:E:accuracy is limited in defining performance}) spills doubt on the \emph{external validity}, what justifies that the measured accuracy is meaningful beyond the measurement context, while the ``target-construct mismatch'' problem (Section~\ref{para:E:accuracy is limited in defining performance}) challenge the \emph{construct validity}, what justifies that the output, and hence the measured accuracy, reflects a relevant concept \citep{raji2021ai, freiesleben2025benchmarking}.
Modern benchmarks even exacerbate these validity concerns, as they are often ignorant about the training procedures, \eg, \citep{glazer2025frontiermathbenchmarkevaluatingadvanced, prandi2025bench, patwatahan2026}. For instance, those benchmarks do not restrict on which data the machine learning models are trained. Hence, it is even harder to justify why a tested accuracy is in fact valid.

\paragraph{Accuracy is limited in defining performance because it only has a comparative character}
When E-accuracy is used as a comparative marker for the ``better'' machine learning system, then it is possible to circumvent some of the concerns on validity. E-accuracy is often a valid relative measurement. On the other hand, E-accuracy is, because of the validity concerns, limited in defining performance, as its absolute scores might not only be uninterpretable, but potentially even vacuous without baseline comparisons.

\subsection{E-accuracy \textcolor{teal}{is poorly suited to} subgroups and individuals}
\label{E:PARTS AND WHOLES}
\textbf{For what is E-accuracy defined?}
The statistical nature of accuracy, being based on aggregates, might spark the question of the role of the individual in accuracy. Does there exist something like ``individual accuracy''. Can a machine learning model be accurate on a single instance? These questions play a surprising role when comparing to the conception of accuracy from a legal perspective \citep{frohnapfel2026using}.

\paragraph{Individual Accuracy is Oxymoronic}
First, the formal definition of accuracy is applicable to singleton sets of instances (\cf \eqref{eq:average accuracy}). Hence, the individual measurement of accuracy is, if the accuracy metric is appropriately defined, possible. One can find an answer to the question whether a machine learning model is accurate on a given instance. The problem of ``individual accuracy'' is not in the measurement, but in the argument whether a machine learning model is accurate on an unseen instance.

By asking the question `` what is the justification that a machine learning method assigns a certain decision, prediction or recommendation to an individual instance?'' we'll argue that individual assignments of machine learning methods are hardly justified and hence individual accuracy is oxymoronic.

Model multiplicity or the ``Rashomon effect'' is the phenomenon that several statistical methods perform accurately on a data set or a data distribution, but assign different predictions to individuals \citep{breiman2001statistical, semenova2022existence}.
Because of modeling decisions, choices of training data, \etc, there exists a ``garden of forking paths'' \citep{simson2025preventing} which results in a set of machine learning models, which are all accurate, but differ on individuals \citep{marx2020predictive, semenova2022existence, black2022model, gorecki2026monoculture}.
Thus, model multiplicity questions the justification of individual assignments of machine learning methods.

The arbitrariness of the individual assignment, as well called predictive multiplicity \citep{marx2020predictive, black2022model}, predictive inconsistency \citep{greene2022forks} or individual arbitrariness \citep{long2023individual}, is not problematic per se. The individual room for variation can help against algorithmic monoculture \citep{creel2022algorithmic,black2022model, gur2025consistently, gorecki2026monoculture} or allow for the inclusion of other design conditions, \eg, such as fairness \citep{greene2022forks, black2022model, black2024legal}.
However, the ``choice among equally accurate models lacks a principled rationale'', hence, ``resulting predictions can lack justification'' \citep{gur2025consistently}.
In particular, the harm of the individual arbitrariness ``is not that the decision is random, but that it is unjustified.'' \citep{gur2025consistently}
Since, the individual assignment is unjustified, its not justified to be accurate either.
It follows that individual arbitrariness renders individual accuracy a non-achievable goal.

Because accuracy is an aggregate notion, it actually implies a multiplicity of accurate model. Imagine there are two machine learning models which only differ in their assignment on a single instance. Such a difference would be impossible to be detected by aggregate quality notions such as accuracy. Hence, model multiplicity is a natural consequence of the statistical nature of accuracy. Only removing the statistical aspect of accuracy, whatever that means, would solve the inherent contradiction in ``individual accuracy''.

\paragraph{Individual Accuracy is Challenged in the Way How Individual Probability is Challenged}
Already Venn observed in 1888 (!) \citep{venn1888logic}, that a statistical statement such as ``penguin $P$ has a probability of $50\% $ of having flue'' is based on placing the penguin within a \emph{reference class}, \ie, all penguins, all penguins of one species, a colony, and deriving the statistical statement on this reference class, \eg, by counting the frequency \citep{reichenbach1949theory}. That entails however the choice of a reference class.
For each choice of reference class the individual probability assignment to the penguin can vary. Thus, an analogous difficulty to model multiplicity for the justification of the individual assignment arises.

That problem returns in more modern approaches to individual\emph{ized} probability assignments \citep{dawid2017individual}. Suppose we argue that a statistical statement on an individual is justified, because the method which produced the statement is ``accurate''\footnote{Most of the approaches based on this justification refer to calibration instead of accuracy in a classical statistical understanding. A probabilistic predictor is calibrated if the averages over the instances for which one specific prediction is given equals the prediction. Calibration can be understood as a special form of abstract statistical accuracy as following --  rather than the predicted label being accurate, it effectively demands that predicted probability is accurate.} on a reference class (\cf i2G in \citep{dawid2017individual}, or in I2A \citep{four2026derr} or \citep{holtgen2024practical}). Then, what justifies the choice of reference class? Or, what justifies the exact method to be used (model multiplicity)? In particular, we end up in circularity. When the individual statistical statement is rationalized based on aggregate accuracy, then accuracy of the individual statement is based upon aggregate accuracy (\cf \citep{hu2025does}). A method is accurate on an individual, because it is accurate on the aggregate. That is an empty statement.

\bibliography{main}


\end{document}